\documentclass[12pt, draftclsnofoot, onecolumn]{IEEEtran}
\usepackage[T1]{fontenc}
\usepackage[utf8]{inputenc}
\usepackage{amsmath} 
\usepackage{amsfonts} 
\usepackage{amssymb} 
\usepackage{amsthm}
\usepackage{color}
\usepackage{url}
\usepackage{caption}
\usepackage{subcaption}
\usepackage[linesnumbered,ruled,vlined]{algorithm2e}
\usepackage{setspace} 
\usepackage{verbatim}
\usepackage{multirow}
\usepackage{graphicx}
\usepackage{comment}

\SetKwInput{KwInput}{Input}                
\SetKwInput{KwOutput}{Output}              
\SetKwInput{Kwinitialize}{Initialization}              

\theoremstyle{remark}

\usepackage{cite}
\renewcommand{\vec}[1]{{\bf{#1}}} 
\newcommand{\vecgreek}[1]{{\boldsymbol{#1}}} 
\newcommand{\tran}{^{\mbox{\scriptsize T}}}
\newcommand{\herm}{^{\mbox{\scriptsize H}}}

\newcommand{\fro}[1]{\Vert #1\Vert_\mathrm{F}^2}
\newcommand{\norm}[1]{\Vert #1\Vert_2}
\DeclareMathOperator{\Tr}{Tr}

\newcommand{\gammab}{\vecgreek{\gamma}}

\newcommand{\mybibliography}{\bibliography{conf_short,jour_short,Bi}}

\begin{document}
\title{Spatial Correlation Aware Compressed Sensing for User Activity Detection and Channel Estimation in Massive MTC}
\author{\IEEEauthorblockN{Hamza Djelouat,~\IEEEmembership{Student Member,~IEEE,} Markus Leinonen,~\IEEEmembership{Member,~IEEE,}  and Markku Juntti,~\IEEEmembership{Fellow,~IEEE}}
\thanks{The authors are with Centre for Wireless Communications -- Radio Technologies, FI-90014, University of Oulu, Finland. e-mail: \{hamza.djelouat,markus.leinonen,markku.juntti\}@oulu.fi.\\
This work has been financially supported in part by the Academy of Finland (grant 319485) and Academy of Finland 6Genesis Flagship (grant 318927). The work of M. Leinonen has also been financially supported in part by Infotech Oulu and the Academy of Finland (grant 323698). H. Djelouat  would like to acknowledge the support of Tauno Tönning Foundation,  Riitta ja Jorma J. Takanen Foundation, and Nokia Foundation.}}
\maketitle

\vspace{-10mm}
\begin{spacing}{1.4}
\begin{abstract}
Grant-free access is considered as a key enabler for massive machine-type communications (mMTC) as it promotes energy-efficiency and small signalling overhead. Due to the sporadic user activity in mMTC, joint user identification and channel estimation (JUICE) is a main challenge. This paper addresses the JUICE in single-cell mMTC with single-antenna users and a multi-antenna base station (BS) under spatially correlated fading channels. In particular, by leveraging the sporadic user activity, we solve the JUICE in a multi measurement vector compressed sensing (CS) framework under two different cases, with and without the knowledge of prior channel distribution information (CDI) at the BS. First, for the case without prior information, we formulate the JUICE as an iterative reweighted $\ell_{2,1}$-norm minimization problem. Second, when the CDI is known to the BS, we exploit the available information and  formulate the JUICE from a Bayesian estimation perspective as a maximum \emph{a posteriori} probability (MAP) estimation problem. For both JUICE formulations, we derive efficient iterative solutions based on the alternating direction method of multipliers (ADMM). The numerical experiments show that the proposed solutions achieve higher channel estimation quality and activity detection accuracy with shorter pilot sequences compared to existing algorithms.
\end{abstract}
\end{spacing} 
\vspace{-5mm}  

\section{Introduction}
Massive machine-type communications  (mMTC)  aim to provide wireless connectivity to billions of low-cost energy-constrained internet of things (IoT) devices \cite{bockelmann2016massive}.  mMTC promote three main features. First,  sporadic transmissions, i.e., only an unknown subset of the IoT devices are active at a given transmission instant. Second, short-packet communications dominated by the  uplink traffic. Third, energy-efficient communication protocols to ensure a long lifespan for the IoT devices, here referred to as user equipments (UEs). As the base station (BS) aims to serve a massive number of energy-constrained devices, channel access management is considered as one of the main  challenges in mMTC \cite{shahab2020grant}.  The conventional channel access protocols, where each UE is assigned a dedicated transmission resource block, are inefficient because many resource blocks are frequently wasted as being pre-assigned to inactive UEs.  Subsequently, alternative schemes have been proposed to provide more efficient channel access protocols. In particular,  \emph{grant-free} multiple-access has been identified as a key enabler  for mMTC \cite{cirik2019toward}.

In the conventional grant-based channel access protocols,  the active UEs first request an access to the channel, and then, the BS allocates a dedicated transmission block to each active UE in a multi-step handover process \cite{shahab2020grant}. Differently,  in grant-free access, the  UEs transmit data as per their needs without going through the  grant-based  access protocols. The main advantage of grant-free access compared to conventional random access is   the reduced signalling overhead and the improved energy-efficiency of the UEs. However, a paramount challenge  in grant-free access is to identify the   \emph{set of active UEs} and to estimate  their \emph{channel state information}  for coherent data detection. We refer to this problem  as  \emph{joint user identification and channel estimation} (JUICE).

The sparse user activity pattern induced by the sporadic transmissions in mMTC motivates the formulation of the JUICE as a \textit{compressed sensing} (CS) \cite{Candes-Romberg-Tao-06,Donoho-06,Haupt-Nowak-06}  problem. Furthermore, as the BS antennas sense the same sparse user activity, the JUICE problem extends to the multiple measurement vector (MMV) CS framework. Sparse support and signal recovery from an MMV setup has been studied extensively in the literature. In a nutshell, the proposed  MMV sparse recovery algorithms can be  categorized into the following classes: 1) greedy algorithms such as simultaneous orthogonal matching pursuit (SOMP)  \cite{tropp2006algorithms}, 2) mixed norm optimization approaches  \cite{steffens2018compact} (and the references therein), 3) iterative methods such as approximate message passing (AMP)\cite{ziniel2012efficient}, and 4) sparse Bayesian learning (SBL) \cite{wipf2007empirical}.

In sparse support and signal recovery algorithms, the prior knowledge on the distributions and the structure of the signals has a profound effect on the recovery performance. For instance, when the signal distribution is  known, algorithms like SBL have shown superior performance compared to mixed-norm minimization \cite{zhang2011sparse}. However, if the signal distribution is unknown and signal statistics are not available,   algorithms based on mixed-norm minimization such as $\ell_{2,1}$-norm minimization present a good choice, since they are invariant to the signal distribution. However, the $\ell_{2,1}$-norm suffers from a bias toward large coefficients in the recovery. Therefore, formulating the sparse recovery as an iterative reweighted $\ell_{1}$-norm \cite{candes2008enhancing} or $\ell_{2,1}$-norm \cite{wipf2010iterative} problem provides a significant improvement compared to their non-reweighted counterparts.

\subsection{Related Work}
A rich line of research has been presented for grant-free access in mMTC. In \cite{chen2018sparse}, Chen {\it et al.}  addressed the user activity detection problem in grant-free mMTC using AMP and derived an analytical performance of the proposed AMP algorithm in both single measurement vector and MMV setups. Liu {\it et al.}  \cite{liu2018massive,liu2018massiveII} extended the analysis of \cite{chen2018sparse} and  conducted an asymptotic performance analysis for activity detection, channel estimation, and achievable rate. Senel and Larsson \cite{senel2018grant} designed a ``non-coherent''  detection scheme for very-short packet transmission by jointly detecting the active users and the transmitted information bits. Ke {\it et al.} \cite{ke2020compressive}  addressed  the JUICE problem in an enhanced mobile broadband  system and proposed a generalized AMP algorithm that exploits the channel sparsity present in both the spatial and the angular domains. 
Yuan {\it et al.} \cite{yuan2019toward} addressed the JUICE problem in a distributed  mMTC system with mixed-analog-to-digital converters under two different user traffic classes.
An SBL approach has been adopted in \cite{zhang2017novel} and a maximum likelihood estimation approach using the measurement covariance matrix has been considered in \cite{chen2019covariance}. Recently, Ying  {\it et al.} \cite{cui2020jointly} presented a model-driven framework for the JUICE by utilizing CS techniques in a  deep learning framework  to jointly design the pilot sequences and detect the active UEs.

In addition to  the sparsity of the activity pattern of the UEs,  the aforementioned algorithms require different degrees of prior information  on the (sparse) signal distribution. For instance, in the AMP-based approaches  \cite{chen2018sparse,liu2018massive,liu2018massiveII,senel2018grant,chen2019covariance}, the BS is assumed to know the distributions and  the large-scale fading coefficients  of channels. The work in \cite{ke2020compressive} relies similarly on the known channel distributions but assumes unknown large-scale fading coefficients, which are estimated via an expectation-maximization approach.

\subsection{Main Contribution}
This paper considers the JUICE problem in single-cell mMTC,  with single-antenna UEs under  spatially correlated multiple-input multiple-output (MIMO)\footnote{In fact, the channels herein are multi-user single-input multiple-output (MU-SIMO) channels. However, we adopt the common ``MIMO'' terminology, which implies that the single-antenna users are the multiple inputs and the  BS antennas are the multiple outputs of the channel.} channels. In particular, we address the JUICE under two different cases, with and without the availability of the \emph{channel distribution information} (CDI) at the BS. First, under unknown CDI, the JUICE is formulated as an iterative reweighted $\ell_{2,1}$-norm minimization with a deterministic regularization penalty that accounts for the sparsity in the user activity. Second, when the CDI is available to the BS, we formulate the JUICE problem from the Bayesian perspective. By using the available knowledge on the CDI and imposing a sparsity-inducing prior on the sporadic activity of the UEs, we formulate the JUICE under a maximum  \emph{a posteriori} probability (MAP)  estimation  framework. For both JUICE  formulations, we derive computationally efficient iterative solutions based on alternating direction method of multipliers (ADMM).

The vast majority of JUICE  works assume that the communications channels are  spatially uncorrelated and often also independent Rayleigh fading. Although this assumption may lead to analytically tractable solutions, it is not always practical as the MIMO  channels are almost always spatially correlated \cite{bjornson2016massive}. Our paper aims to bridge this gap by addressing spatially correlated channels, which have not been widely studied in the context of JUICE in mMTC. In fact, incorporating the spatial correlation structure in the design of a JUICE solution is crucial, because the performance of JUICE solutions designed for uncorrelated channels may be sensitive to the correlation structures faced in practical scenarios \cite{cui2020jointly}.   Recently, Chen  {\it et al.} \cite{cheng2020orthogonal} presented an orthogonal AMP algorithm to exploit both the spatial  channel correlation in mMTC systems.

The main \textbf{contributions} of our paper can be summarized as follows:
\begin{spacing}{1.4}
\begin{itemize}
\item We address the JUICE problem in spatially correlated MIMO channels to provide a realistic assessment of the performance of the proposed JUICE solutions. Although precise knowledge of the CDI may be challenging in some practical applications, the results provide channel estimation performance benchmark for system design.
\item When the BS has limited knowledge on the data structure, i.e., only the sparse behaviour of users activity is taken into consideration, we exploit the benefits of reweighting strategies in CS and formulate the JUICE as a reweighted iterative $\ell_{2,1}$-norm optimization problem. Reweighted $\ell_{2,1}$-norm minimization has not been used for JUICE problems earlier.
\item When the CDI is known, we fully exploit the available information and  propose a novel JUICE  formulation from the Bayesian perspective. The proposed formulation relaxes non-convex Bayesian MAP estimation to convex regularization-based optimization. In particular, the CDI knowledge is incorporated via the Mahalanobis distance measure.  
\item For each JUICE formulation, we  use a specific variable splitting strategy that allows to derive an exact  ADMM solution. The proposed approach decouples the JUICE problem into a set of convex sub-problems, each admitting a computationally efficient closed-form solution that can be computed efficiently via a simple analytical formula.
 \item  We show empirically that the proposed algorithms enhance the accuracy of user activity detection and  channel estimation quality. In particular, for  predefined requirements,  the proposed approaches achieve the same performance as baseline MMV JUICE solutions even when using significantly  smaller signalling overhead. 
\end{itemize}
\end{spacing}

This paper is in line with our recent work \cite{Djelouat2020Joint,Djelouat-Leinonen-Juntti-21-icassp} where we addressed the JUICE under spatially correlated highly directive channels. In \cite{Djelouat2020Joint,Djelouat-Leinonen-Juntti-21-icassp}, the JUICE  was formulated as a mixed-norm minimization problem, augmented by a deterministic penalty that exploits the second-order statistics of the channels. In this paper, we further leverage the available knowledge on the entire CDI and treat the JUICE problem under a more rigorous, Bayesian framework.

\textit{Organization:} The rest of the paper is organized as follows. Section \ref{sec::system}
presents the system model and the canonical JUICE problem formulation. Section \ref{sec::itertative} addresses the JUICE with unknown CDI. Section \ref{sec::MAP-JUICE} derives the Bayesian formulation for the MAP-based JUICE which exploits the prior knowledge on the CDI.  Simulation results are provided in Section \ref{Result}, and Section \ref{conclusion} concludes the paper.

\textit{Notations:} Throughout this paper, we use boldface uppercase letters $(\vec{A})$ to denote matrices, boldface lowercase letters $(\vec{a})$ for vectors, and calligraphy letters $(\mathcal{S})$ to denote sets. The $i$th column of matrix $\vec{X}$ is denoted by $\vec{x}_i$. The transpose, the Hermitian, and the conjugate  of a
matrix are denoted as $(\cdot)\tran$, $(\cdot)\herm$, and $(\cdot)^*$, respectively. $\vec{0}$ and $\vec{1}$ are vectors of all entries zero and one, respectively. The $\ell_2$-norm and the Frobenius norm are  denoted as $\|\cdot\|_2$ and $\|\cdot\|_{\mathrm{F}}$, respectively. $\|\vec{a}\|_0$ counts the number of non-zero entries of vector $\vec{a}$. $1(a)$ is an indicator function that takes the value 1 if $a\neq0$, and 0 otherwise.  $\otimes$  denotes the Kronecker product and $ \mathrm{vec}(\cdot)$ denotes the operation of column-wise stacking of a matrix.

\section{System Model and Problem Formulation}
\label{sec::system}
\subsection{System Model}
\begin{figure}[t!]
    \centering
     \begin{subfigure}{0.32\textwidth}
    \includegraphics[scale=.39]{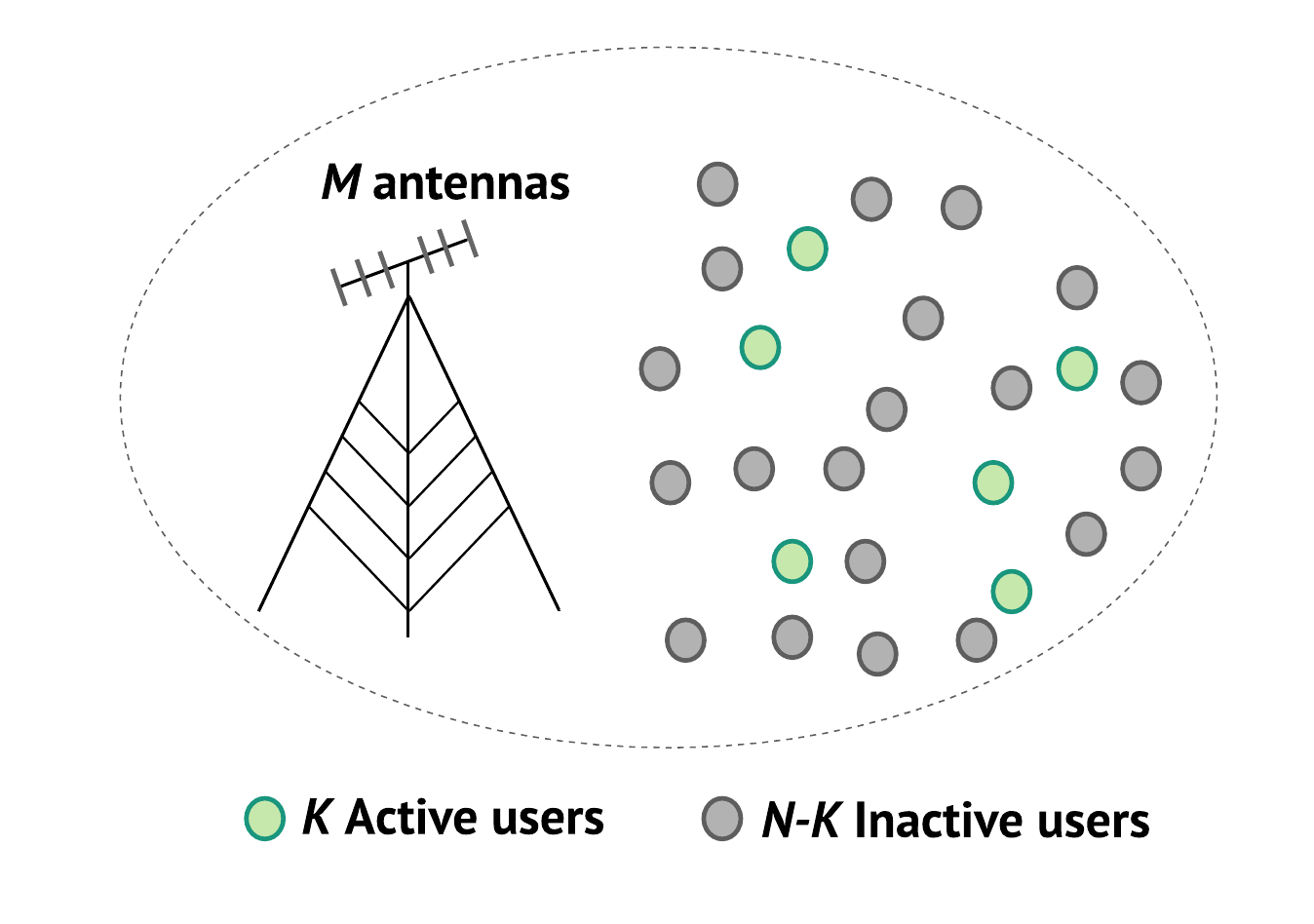}
    \caption{}
    \label{fig:mMTC}
\end{subfigure}
   \hfill      
   \begin{subfigure}{0.32\textwidth}
  \centering
     \includegraphics[scale=.39]{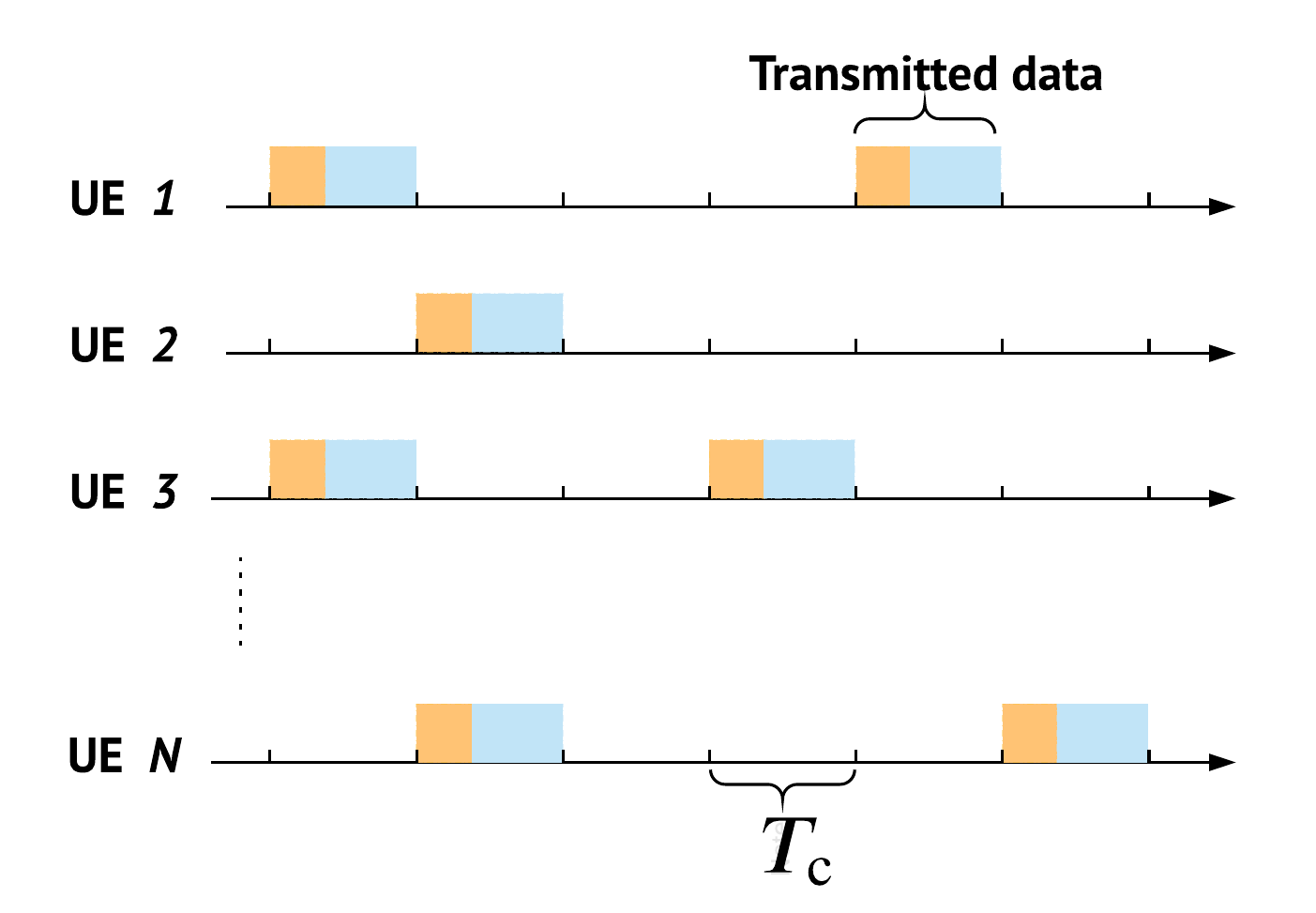}
    \caption{}
     \label{fig:sporadic}
\end{subfigure}  
\hfill
    \begin{subfigure}{0.32\textwidth}
  \centering
     \includegraphics[scale=.39]{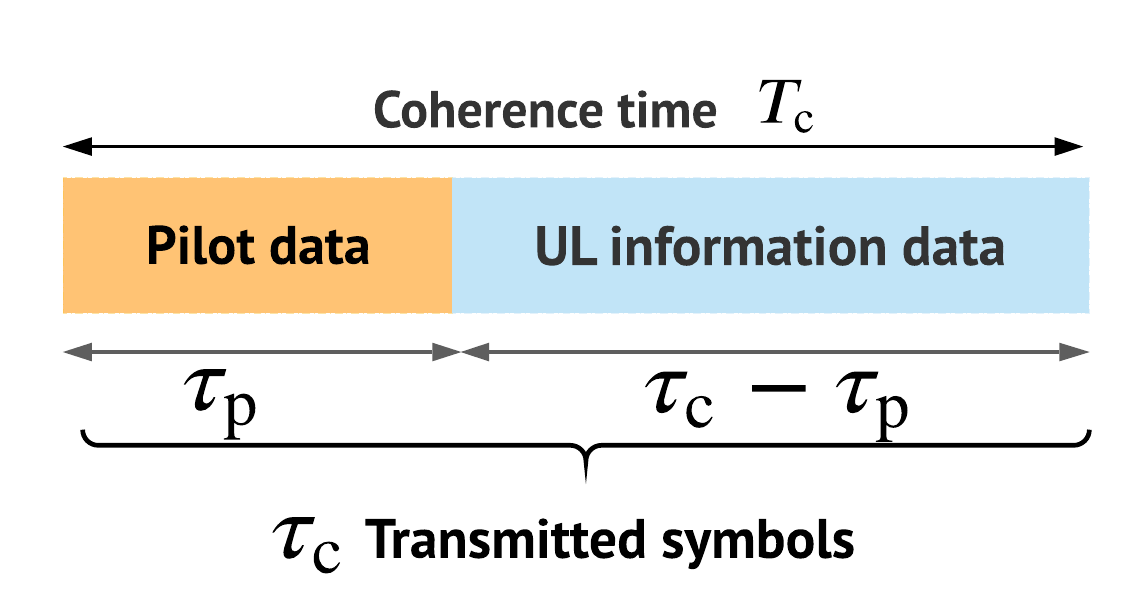}
     \caption{}
     \label{fig:T_c}
\end{subfigure}
\vspace{-3mm}
\caption{Illustration of a typical mMTC scenario:  (a) an mMTC uplink system with $K$ active UEs and $N-K$ inactive UEs, (b) sporadic transmission, (c) division of a coherence interval $T_{\mathrm{c}}$.}
\label{fig::MTCsystem}\vspace{-6mm}
\end{figure}

We consider a single-cell uplink mMTC system, as depicted in Fig.\ \ref{fig::MTCsystem}(a). The cell consists of a set $\mathcal{N}=\{1,\ldots,N\}$  uniformly distributed single-antenna UEs communicating with a  BS  equipped with a uniform linear array (ULA) containing $M$ antennas.  

We consider a block fading channel response over each coherence period $T_{\mathrm{c}}$. Furthermore, to model the propagation channels between  the UEs and the BS, we consider a \emph{local scattering model}, which is suitable for multi-antenna channel  modelling as it can capture some  key characteristics of the typical MIMO channels \cite[Sect.~2.6]{massivemimobook}. In the local scattering channel model, the BS is considered to be located in an elevated position and thus, it has no scatterers in its near-field, whereas the UEs are surrounded by rich scattering environment. The channel response vector from each UE $ i\in \mathcal{N}$ is modelled as the superposition of $P_i$ physical signal paths, each reaching the BS as a plane wave.   Accordingly, the channel response vector  between the $i$th UE and the BS, denoted as  ${\vec{h}_{i}\in\mathbb{C}^{M}}$, is modelled as
\begin{equation}
    \vec{h}_i=\frac{1}{\sqrt{P_i}}\sum_{p=1}^{P_i}g_{i,p}\vec{a}(\psi_{i,p}),
    \label{eq::ch}
\end{equation}
where $g_{i,p} \in \mathbb{C}$ accounts for the gain and the phase-rotation of the $p$th propagation path, $\psi_{i,p}$ is the angle of arrival (AoA) of the $p$th path, and $\vec{a}(\psi_{i,p}) \in \mathbb{C}^{M}$ is the steering vector of the ULA,  defined as $\vec{a}(\psi_{i,p})=[1, e^{-j2\pi\Delta_\mathrm{r} \cos(\psi_{i,p})}, \ldots, e^{-j2\pi(M-1)\Delta_\mathrm{r} \cos(\psi_{i,p})}]\tran$, where $\Delta_\mathrm{r}$ denotes the normalized spacing between the adjacent BS antennas. We consider that  $\psi_{i,p}=\bar{\psi}_{i} +\zeta_{i,p}$, where  $\bar{\psi}_{i} \in [-\pi/2,\pi/2]$ represents the (deterministic) incident angle between the $i$th user and the BS, and $\zeta_{i,p}$ denotes a (random) deviation from the incident angle with angular standard deviation $\sigma_{\psi}$. We assume that each $\zeta_{i,p}$  follows a Gaussian distribution $\zeta_{i,p}\sim\mathcal{CN}(0,\,\sigma_{\psi}^{2})$ \cite[Sect.~2.6]{massivemimobook}.

The propagation channel between each UE and the BS is often considered to follow a complex Gaussian distribution. More specifically, by utilizing the valid  assumption that the  number of scatterers around each UE is very large in practice and invoking the central  limit theorem, the channel vector $\vec{h}_i$ in \eqref{eq::ch}  can be modelled as a complex Gaussian random variable with zero mean and covariance matrix $\vec{R}_i=\mathbb{E}[\vec{h}_i\vec{h}_i\herm] \in \mathbb{C}^{M \times M}$ \cite[Sect.~2.6]{massivemimobook}, i.e., 
\begin{equation}
    \vec{h}_i \sim  \mathcal{CN}(0,\vec{R}_i),\;\; P_i  \rightarrow \infty,~ \forall i\in \mathcal{N}\label{eq::CN}.
\end{equation} 
The channel realizations $\vec{h}_i$  are independent between different coherence intervals $T_\mathrm{c}$. We consider UEs with low mobility, which is justified in the context of mMTC \cite{laya2013random}. Hence, we adopt the common assumption that the channels are  wide-sense stationary \cite{Li-etal15}. Thus,  the  set of channel covariance matrices $\{\vec{R}_i\}_{i=1}^N$  vary in a slower timescale  compared to the channel realizations \cite{sanguinetti2019towards}. Accordingly,   $\{\vec{R}_i\}_{i=1}^N$ are assumed to remain fixed for $\tau_\mathrm{s}$ coherence intervals, where $\tau_\mathrm{s}$ can be on the order of thousands \cite{bjornson2016massive}. We note that this assumption can be challenging in mMTC where some UEs are inactive for a longer period. Therefore, we will elaborate further on this issue in Section \ref{sec:CDI}.

Due to the sporadic activity pattern of mMTC, only ${K<<N}$ UEs are active at each coherence interval  $T_{\mathrm{c}}$, whereas the remaining $N-K$ are inactive. This is depicted in Fig.\ \ref{fig::MTCsystem}(b). In order to deploy a  grant-free multiple access scheme, we assume that all the UEs and the BS are synchronized. In addition, each  coherence interval  $T_{\mathrm{c}}$ permits transmitting $\tau_{\mathrm{c}}$ symbols and is divided into two phases, as shown in Fig.\ \ref{fig::MTCsystem}(c). In the first phase, each active UE transmits its $\tau_{\mathrm{p}}$-length pilot sequence to  the BS. In the  second phase, using the remaining $\tau_{\mathrm{c}}-\tau_{\mathrm{p}}$ symbols,  the active UEs send their information data to the BS. During each  $T_{\mathrm{c}}$, the BS uses the transmitted pilot sequences from the first phase to identify the set of active UEs and  estimate their corresponding  channels in order to decode the information data transmitted at the second phase.

Regarding the channel estimation phase, the BS assigns to each UE $ i\in \mathcal{N}$  a unique unit-norm pilot sequence $\vecgreek{\phi}_i \in \mathbb{C}^{\tau_{\mathrm{p}}}$. Due to the potentially large number of UEs, the UEs cannot be assigned orthogonal pilot sequences, because it would require a pilot length of the same order as the number of UEs. Therefore, the BS assigns a set of non-orthogonal pilots which can be generated, for instance, from  an independent  identically distributed (i.i.d.) Gaussian or i.i.d.\  Bernoulli distribution. Herein we consider pilot sequences generated from a complex symmetric Bernoulli distribution.  This approach would drive the probability of pilot collision, i.e., assigning the same pilot to two distinct UEs, to be negligible \cite{senel2018grant}.

Furthermore,  to  mitigate the channel gain differences between the UEs, a power control policy is deployed such that each UE $ i\in \mathcal{N}$  transmits with a power $p^{\mathrm{UL}}_i$ that is inversely proportional to the average channel gain \cite{senel2018grant,bjornson2016massive}. 
Accordingly, the received signal associated with the transmitted pilots at the BS, denoted by $\vec{Y} \in \mathbb{C}^{\tau_{\mathrm{p}}\times M}$,  is given by
  \begin{equation}
\label{eq::Y}
 \vec{Y}=\sum_{i=1}^{N}\gamma_i  \sqrt{p^{\mathrm{UL}}_i}  \vecgreek{\phi}_i\vec{h}_i\tran+\vec{W},
\end{equation}
where $\vec{W} \!\!\in\!\mathbb{C}^{\tau_{\mathrm{p}}\times M}$ is an additive white Gaussian noise with independent an i.i.d.\ elements as $\mathcal{CN}(0,\,\sigma^{2})$, and $\gamma_i\!\in\mathbb{B}$ is the $i$th element of the binary user activity indicator vector $\vecgreek{\gamma}=[\gamma_1,\gamma_2,\ldots,\gamma_N]\tran$, defined as 
\begin{spacing}{1.1}
\begin{equation}
\gamma_i =
\begin{cases}
1, & i \in \mathcal{S}\\ 
0 , & \text{otherwise},
\end{cases}\qquad \forall i \in \mathcal{N} 
\end{equation}
\end{spacing}
\noindent where $\mathcal{S}\subseteq\{1,\ldots,N\}$,  ${|\mathcal{S}|=K}$, is the set of active users. We assume that besides not knowing which users are active at a given time, the BS does not either know the  activity level $\frac{K}{N}$.

Let us define the effective channel of user ${i\in\mathcal{N}}$ as $\vec{x}_i=\gamma_i \sqrt{p^{\mathrm{UL}}_i}\vec{h}_i$, and subsequently, the effective channel matrix  as $\vec{X}=[\vec{x}_1,\ldots,\vec{x}_{N}] \in \mathbb{C}^{M\times N}$. The pilot sequence matrix is defined as $\vec{\Phi}=[\vecgreek{\phi}_1,\ldots,\vecgreek{\phi}_N] \in \mathbb{C}^{\tau_\mathrm{p}\times N}$. Accordingly, we can rewrite the received signal associated with the pilots in  \eqref{eq::Y} as
\begin{equation}
     \vec{Y}= \vec{\Phi} \vec{X}\tran+ \vec{W}. 
     \label{eq::CS}
\end{equation}

\subsection{Problem Formulation}
The columns of effective channel matrix $\vec{X}$ corresponding to the inactive users are zeros, thus, $\vec{X}\tran$ is  a \emph{row-sparse} matrix; it contains only $K$ non-zero rows. The objective of JUICE is to jointly identify and estimate  the non-zero elements of effective channel matrix  $\vec{X}$. Thus, JUICE can be modelled as joint support and signal recovery from an MMV setup. Subsequently, the  canonical form of the JUICE can be presented as
\begin{equation}
       \min _{\vec{X} }\frac{1}{2}\|\vec{\Phi}\vec{X}\tran-\vec{Y}\|_{\mathrm{F}}^2 + \beta_1 \|\vec{X}\tran\|_{2,0},
     \label{eq::l_0}
\end{equation}
where $\|\vec{X}\tran\|_{2,0}=\sum_{i=1}^{N}1(\|\vec{x}_i\|_2)$ is the sparsifying regularizer and $\beta_1$  controls the trade-off between the emphasis on the measurement consistency term and the sparsity-promoting term. However, the $\ell_0$-``norm'' (with slight abuse of terminology regarding a norm) minimization is  an intractable combinatorial NP-hard  problem. Therefore, several algorithms have been presented in the literature to relax the optimization problem \eqref{eq::l_0}. The existing  algorithms  can be categorized based on their required prior information on the signal. For instance, while AMP and SBL require   prior information on the distributions of a sparse signal, mixed-norm minimization and most of the greedy algorithms operate based on the mere fact that the signal has a sparse structure.

In this paper, we cover both cases, i.e., JUICE with and without  prior knowledge on the CDI. First, when there is no prior knowledge on the channel, we formulate  the JUICE as  an iterative reweighted   $\ell_{2,1}$-norm optimization  problem in Section \ref{sec::itertative}. Second, in Section \ref{sec::MAP-JUICE}, we assume that the BS has prior knowledge on the CDI, and we formulate the JUICE as  a MAP estimation problem. For both these JUICE frameworks, we will derive  a computationally efficient ADMM method to solve the formulated optimization problem. Each ADMM algorithm solves a relaxed version of the involved problem iteratively, and in particular, provides a closed-form solution to each sub-problem included in the optimization process. 

\section{JUICE via Reweighted $\ell_{2,1}$-Norm Minimization}
 \label{sec::itertative}

Without the CDI, the $\ell_{2,1}$-norm penalty is commonly used to relax  the $\ell_{2,0}$-norm penalty in the JUICE formulation in \eqref{eq::l_0} as
\begin{equation}\label{eq::l2,1}
   \min _{\vec{X}}\frac{1}{2}\Vert \vec{\Phi} \vec{X}\tran - \vec{Y} \Vert _{\mathrm{F}}^2 + \beta_1 \Vert \vec{X}\tran\Vert_{2,1}.  
  \end{equation}
Nevertheless, unlike the democratic $\ell_0$-norm  which penalizes the  non-zero coefficients equally, $\ell_1$-norm is biased toward larger magnitudes, i.e.,  coefficients with a large magnitude are penalized more heavily than smaller ones. 
Therefore, striving for enhanced sparsity recovery, we use the \emph{log-sum} penalty \cite{candes2008enhancing} to relax the $\ell_{0}$-norm in \eqref{eq::l_0} as 
\begin{equation}
\begin{array}{ll}
& \displaystyle\min_{\vec{X},\vec{u} }\displaystyle\frac{1}{2}\|\vec{\Phi} \vec{X}\tran - \vec{Y}\|_{\mathrm{F}}^2+\beta_1 \sum_{i=1}^N \log(u_i+\epsilon_0) \;\;\;  \mbox{s.t.} \;\;\;\norm{\vec{x}_i} \leq u_i ,~ \forall i \in \mathcal{N},
 \end{array}
   \label{eq::log-sum}
\end{equation}
where $\vec{u}=[u_1,u_2,\ldots,u_N]\tran$ is a vector of auxiliary optimization variables and $\epsilon_0$ is a small positive stability parameter. The log-sum penalty resembles most closely  the $\ell_{2,0}$-norm penalty when $\epsilon_{0} \!\rightarrow 0$. However, a practical, numerically robust choice is to set $\epsilon_0$  to be slightly less than the expected norm of the non-zero rows in $\vec{X}\tran$ \cite{candes2008enhancing}. 

As the objective function in \eqref{eq::log-sum} is a sum of a convex and a concave functions, it is not convex in general. Therefore, by applying a majorization-minimization (MM) approximation, \eqref{eq::log-sum} can be solved as the following \emph{iterative reweighted $\ell_{2,1}$-norm} minimization problem
\begin{equation}\label{eq::MM_x}
\vec{X}^{(l+1)}=\displaystyle\min_{\vec{X}}\displaystyle  \frac{1}{2} \big\| \vec{\Phi}\vec{X}\tran - \vec{Y} \big\|_{\mathrm{F}}^2 +\sum_{i=1}^{N}\beta_{1} g_{i}^{(l)}\|\vec{x}_{i}\|_{2},
\end{equation}
with the weights set at iteration $(l)$ as
\begin{equation}\label{eq::g}
   g_{i}^{(l)} = (\epsilon_0+\| \vec{x}_{i}^{(l)}\|_{2})^{-1}, \forall i \in \mathcal{N}.
\end{equation}

\subsection{IRW-ADMM Solution}
The optimization problem \eqref{eq::MM_x} is convex and can
be solved optimally using standard convex optimization
techniques. However, as mMTC systems may grow large, the standard techniques can suffer from high computational complexity.
As a remedy, we utilize  ADMM \cite{boyd2011distributed} to solve \eqref{eq::MM_x} iteratively in a computationally efficient manner at each MM iteration $(l)$. 

ADMM  has been widely used to provide computationally efficient solutions to sparse signal recovery problems \cite{lu2011fast}.  Apart from signal reconstruction, ADMM has also been  utilized in the context of activity detection in mMTC   \cite{cirik2017multi,cui2020jointly}. Cirik {\it et al.} \cite{cirik2017multi} proposed an ADMM-based solution to multi-user support and signal detection in an SMV model,  where they  incorporate  prior knowledge on the signal recovered from the previous transmission instants. In addition, Ying {\it et al.} \cite{cui2020jointly} proposed an approximation step in ADMM similar to \cite{lu2011fast}, but they solved the sub-problems through a model-driven deep learning decoder.

In contrast to the approximate solutions to  problem \eqref{eq::MM_x}  provided in \cite{cui2020jointly,lu2011fast},  we solve  \eqref{eq::MM_x} \emph{exactly} by adopting a  variable splitting strategy different to \cite{cui2020jointly,lu2011fast}. More precisely, the  proposed splitting technique decomposes the objective function in \eqref{eq::MM_x} into two separable convex functions that can be solved efficiently via  simple analytical formulas. In particular, we derive a set of update rules to solve \eqref{eq::MM_x} iteratively in a sequential fashion over multiple convex sub-problems, where each sub-problem admits \emph{a closed-form solution}, as we will show next.

By introducing a splitting variable  ${\vec{Z}\in \mathbb{C}^{M\times N}}$,  i.e., a copy of optimization variable $\vec{X}$, we decompose the objective function in \eqref{eq::MM_x} into two separate functions: a quadratic function on the measurement fidelity over $\vec{Z}$ and a reweighted $\ell_{2,1}$-norm penalty over $\vec{X}$. Subsequently, 
we rewrite the optimization problem  \eqref{eq::MM_x} as
\begin{equation}
\begin{array}{ll}
       (\vec{X}^{(l+1)},\vec{Z}^{(l+1)})=\displaystyle\min_{\vec{X},\vec{Z}}\displaystyle\frac{1}{2}\|\vec{\Phi}\vec{Z}\tran-\vec{Y}\|_{\mathrm{F}}^2 +\sum_{i=1}^{N}\beta_{1} g_{i}^{(l)}\|\vec{x}_{i}\|_{2},  \;\;\;\; \mbox{s.t.} \;\;\vec{X}=\vec{Z}.
\end{array}
\label{eq::IRW_problem}
\end{equation}

Next, we write the  augmented Lagrangian of \eqref{eq::IRW_problem}  as follows
 \begin{equation}
 \begin{array}{ll}
 \mathcal{L}(\vec{X},\vec{Z},\vec{\Lambda})=
 \displaystyle  \frac{1}{2}
 \| \vec{\Phi} \vec{Z}\tran -\vec{Y}\|_{\mathrm{F}}^2 +\beta_1\sum_{i=1}^N g_i^{(l)} \norm{\vec{x}_i}  +\displaystyle\frac{\rho}{2}\|\vec{X}-\vec{Z}+\displaystyle\frac{\vec{\Lambda}}{\rho}\|_{\mathrm{F}}^2-\displaystyle\frac{\fro{\vec{\Lambda}}}{2\rho},
 \end{array}
\label{eq::Lagrange_ADMM}
\end{equation}
where  $\vec{\Lambda}=[\vecgreek{\lambda}_1,\ldots,\vecgreek{\lambda}_N] \in \mathbb{C}^{M\times N}$ denotes the  dual variable matrix containing the ADMM dual variables $\{\vecgreek{\lambda}_i\}_{i=1}^N$, and $\rho$ is a positive parameter for adjusting the convergence  of the ADMM.
 
The ADMM  solves an optimization problem through sequential 
phases over the primal variables   followed by the method of
multipliers  to update  the dual variables  \cite{boyd2011distributed}. Therefore, by applying the ADMM to the optimization  problem \eqref{eq::MM_x},
we first minimize   \eqref{eq::Lagrange_ADMM}  over  the primal variable  $\vec{Z}$
with $(\vec{X},\vec{\Lambda})$ fixed, followed by minimization over the primal variable $\vec{X}$  with $(\vec{Z},\vec{\Lambda})$ fixed. Finally, the ADMM updates the dual variable matrix $\vec{\Lambda}$ using the most recent updates of $(\vec{X},\vec{Z})$. Thus, the ADMM for \eqref{eq::MM_x} consists of the following three steps:
\begin{equation}
\vec{Z}^{(k+1)}:=\displaystyle\min_{\vec{Z}}\mathcal{L}(\vec{X}^{(k)},\vec{Z},\vec{\Lambda}^{(k)}):=\displaystyle\min_{\vec{Z}}\frac{1}{2}\| \vec{\Phi}\vec{Z}\tran -\vec{Y}\|_{\mathrm{F}}^2+ \frac{\rho}{2} \Vert   \vec{X}^{(k)} - \vec{Z} +\frac{1}{\rho}\vec{\Lambda}^{(k)} \|_{\mathrm{F}}^2
\label{eq::z(k+1)}
\end{equation}\vspace{-5mm}
\begin{equation}
\vec{X}^{(k+1)}\;\;:=\displaystyle\min_{\vec{X}}\mathcal{L}(\vec{X},\vec{Z}^{(k+1)},\vec{\Lambda}^{(k)})\;\;\;\hspace{-3mm}:=\displaystyle\min_{\vec{X}}  \sum_{i=1}^{N} \beta_1 g_i^{(l)} \Vert  \vec{x}_i\|_2   +\frac{\rho}{2} \|\vec{X}-\vec{Z}^{(k+1)}+ \frac{1}{\rho}\vec{\Lambda}^{(k)} \|_{\mathrm{F}}^2
\label{eq::x(k+1)}
\end{equation}\vspace{-5mm}
\begin{equation}
\vec{\Lambda}^{(k+1)} := \vec{\Lambda}^{(k)}+\rho\big(\vec{X}^{(k+1)}  -\vec{Z}^{(k+1)} \big),
\label{eq::lambda(k+1)}
\end{equation}
where the superscript $(k)$ denotes the ADMM iteration index\footnote{For brevity, the dependency of the ADMM variables (e.g., $\vec{X}$, $\vec{Z}$, and $\vec{\Lambda}$) on the MM iteration index $(l)$ is omitted throughout the paper.}. 
The derivations of the ADMM steps \eqref{eq::z(k+1)} and \eqref{eq::x(k+1)} are detailed  below. 

\paragraph{$\vec{Z}$-update} ADMM updates the primal variable $\vec{Z}$ by solving  the convex optimization problem \eqref{eq::z(k+1)}. Thus,  $\vec{Z}^{(k+1)}$ is obtained by setting the gradient of the objective function in \eqref{eq::z(k+1)} with respect to $\vec{Z}$ to zero, resulting in
\begin{equation}
    \vec{Z}^{(k+1)}=\big(\rho \vec{X}^{(k)} +\vec{\Lambda}^{(k)}+\vec{Y}\tran\vec{\Phi}^*\big) \big( \vec{\Phi}\tran \vec{\Phi}^*+\rho \vec{I}_N\big)^{-1}.\label{eq::Z+}
\end{equation}
Note that the matrix inversion $\big( \vec{\Phi}^*\vec{\Phi}\tran+\rho \vec{I}_N\big)^{-1}$ and the product $\vec{Y}\tran\vec{\Phi}^*$ need to be computed only once, thus, they can be stored, reducing the overall algorithm complexity.
\paragraph{$\vec{X}$-update} The optimization problem  \eqref{eq::x(k+1)} can be decomposed into $N$ sub-problems as  
\begin{equation}
  \vec{x}_i^{(k+1)}:=\min_{\vec{x}_i} \displaystyle\frac{\beta_1 g_i^{(l)}}{\rho}\|  \vec{x}_i\|_2   +\frac{1}{2} \|\vec{x}_i-\vec{c}_i^{(k)}\|_2^2, \;\;\; \forall i \in \mathcal{N},
  \label{eq::l1}
\end{equation}
where $\vec{c}_i^{(k)}=\vec{z}_i^{(k+1)}- \dfrac{1}{\rho}\vecgreek{\lambda}_i^{(k)}$ and $\vecgreek{\lambda}_i^{(k)}$ is the $i$th column of $\vec{\Lambda}^{(k)}$.  The problem in \eqref{eq::l1} admits a closed-form  solution given by the soft thresholding operator \cite{goldstein2014field} as
\begin{equation}
    \vec{x}_i^{(k+1)}= \frac{\max{\Big\{0,\|\vec{c}_i^{(k)} \|_2-\frac{\beta_1 g_i^{(l)}}{\rho}\Big\}}}{\|\vec{c}_i^{(k)} \|_2}\vec{c}_i^{(k)},\quad \forall i \in \mathcal{N}.
    \label{eq::prox}
\end{equation}

Finally, the dual variable update $\vec{\Lambda}^{(k+1)}$ is performed using \eqref{eq::lambda(k+1)}. 
\subsection{Algorithm Implementation}
The details for the  proposed  iterative reweighted ADMM (IRW-ADMM) algorithm  to solve the problem \eqref{eq::MM_x}  are summarized in Algorithm 1. As one stopping criterion, Algorithm 1  is  run until the $\vec{X}$-update is converged, measured as $\fro{\vec{X}^{(k)}-\vec{X}^{(k-1)}}<\epsilon$ with a predefined
tolerance parameter $\epsilon>0$, or until a maximum number of iterations  $l_{\mathrm{max}}k_{\mathrm{max}}$ is reached, where  $l_{\mathrm{max}}$ denotes the maximum number of  iterations in the MM loop  and $k_{\mathrm{max}}$ denotes the maximum number of iterations in the ADMM loop. Note that if the weight vector  is fixed to $\vec{g}^{(l)}=\vec{1},~l=1,2,\ldots$, Algorithm 1 provides the ADMM solution for optimization problem \eqref{eq::l2,1}, which we term  ADMM henceforth.

\begin{algorithm}[h]
\DontPrintSemicolon
   \KwInput{1) Pilot matrix $\vec{\Phi}$, 2) parameters $\beta_1,\rho,\epsilon_0,\epsilon,l_{\mathrm{max}},k_{\mathrm{max}}$}
  \KwOutput{$\hat{\vec{X}}$}
 \Kwinitialize{1) $\vec{X}^{(0)},\vec{V}^{(0)},\vec{Z}^{(0)},\vecgreek{\Lambda}^{(0)}, k=1$, $l=1$, and 2) $\big( \vec{\Phi}\tran\vec{\Phi}^*+\rho \vec{I}_N\big)^{-1}$ }
 Receive $\vec{Y}$ at the BS, and compute and store  $\vec{Y}\tran\vec{\Phi}^*$\\
   \While{$l<l_{\mathrm{max}}$  }
   {
   \While{$k<k_{\mathrm{max}}$ $\mathrm{or}$ $\| \vec{X}^{(k)}-\vec{X}^{(k-1)} \|^2_{\mathrm{F}}<\epsilon$}
   {
Update $\vec{Z}^{(k+1)}$ using  \eqref{eq::Z+}\;
Update $\vec{X}^{(k+1)}$ using  \eqref{eq::prox}\;
 $\vecgreek{\Lambda}^{(k+1)}=  \vecgreek{\Lambda}^{(k)}+\rho\big(   \vec{X}^{(k+1)}-\vec{Z}^{(k+1)} \big)$\;
$k\leftarrow{k+1}$\;
 }
   $\vec{X}^{(l)}\leftarrow \vec{X}^{(k+1)}$\;
  $ g_{i}^{(l)} = (\epsilon_0+\| \vec{x}_{i}^{(l)}\|_{2})^{-1}, i\in\mathcal{N}$ \;
$l\leftarrow{l+1}$\;
}
\caption{IRW-ADMM}
\end{algorithm}
\section{Spatial Correlation Aware JUICE via Bayesian Estimation}
\label{sec::MAP-JUICE}
In this section, we propose a Bayesian formulation for JUICE  when the CDI is available at the BS. We formulate the JUICE as   MAP estimation and  derive a computationally efficient ADMM solution for a relaxed version of the MAP problem. 

\subsection{MAP Estimation} 
The JUICE formulation presented in Section \ref{sec::itertative} as an iterative reweighted $\ell_{2,1}$-norm minimization (problem \eqref{eq::MM_x})  can be viewed as a joint support and signal recovery problem with a deterministic sparsity regularization. Such formulation presents a robust approach as it is invariant to the channel statistics, making it suitable for a broad range of channel distributions. However,  the optimization problem  \eqref{eq::MM_x} omits any available  side information on the CDI.  Alternatively,  if the CDI is available,  the JUICE problem can be formulated in a Bayesian framework to account for the fact that each unknown channel to be estimated is a realization of a random variable (vector) with the known distribution. A Bayesian sparse signal recovery framework has great potential in providing certain advantages over deterministic formulations \cite{babacan2009bayesian}. 

Developing a JUICE solution from a Bayesian perspective is enabled by: 1) the fact that the  propagation channels $\vec{h}_i$, $i \in \mathcal{N}$, are modeled by  Gaussian distributions as shown in \eqref{eq::CN}, and 2)
the relatively slowly changing covariance matrices $\{\vec{R}_i\}_{i=1}^N$
which can be estimated with high accuracy. In the rest of the paper, we consider the common assumption that  $\{\vec{R}_i\}_{i=1}^N$ are  known to the BS \cite{Li-etal15}. The acquisition of CDI knowledge is further elaborated in Section~\ref{sec:CDI}.

Next,  we utilize the prior information on the CDI and  derive a Bayesian formulation  for the JUICE problem. The JUICE performs two tasks in a joint fashion: 1) identification of the support of the user activity indicator vector $\gammab$, and 2) estimation of the effective channel matrix $\vec{X}$, relying on the current estimate of $\gammab$. The JUICE formulation in  \eqref{eq::MM_x}  applies  a deterministic penalty  that accounts for the row-sparsity of  $\vec{X}\tran$ which inherently captures the sparsity in $\gammab$. However, in the Bayesian modelling, we treat the two variables to be estimated, $\gammab$ and  $\vec{X}$, as unknown quantities with such prior distributions that best model our knowledge on their true distributions, that is: 1) the sparse distribution of the user activity indicator vector   $\gammab$, and 2) the effective channel $\vec{x}_i,~\forall i\in \mathcal{N}$, which is a random vector consisting of a multiplication of $\gamma_i$ and the complex Gaussian random vector  $\vec{h}_i$  (i.e., $\vec{x}_i=\sqrt{p^{\mathrm{UL}}_i} \gamma_i \vec{h}_i$).

We derive   joint MAP estimates $\{\hat{\vec{X}},\hat{\gammab}\}$ by making an explicit use of the prior knowledge on the fact that the propagation channels between the UEs and the BS follow complex Gaussian distributions given in  \eqref{eq::CN}, under the assumption that the BS knows the estimates of the second-order statistics of the channels, i.e., the matrices $\{\hat{\vec{R}}_i\}_{i=1}^N$. To this end,  the joint MAP estimates $\{\hat{\vec{X}},\hat{\gammab}\}$ with respect to the posterior density given the measurement matrix $\vec{Y}$ is given by
\begin{equation}
\begin{array}{ll}\label{eq:map_x_gamma}
\{\hat{\vec{X}},\hat{\gammab}\}&\hspace{-3mm}=\underset{\vec{X},\gammab}{\max}~\displaystyle p(\vec{X},\gammab|\vec{Y})\\
&\hspace{-3mm}=\underset{\vec{X},\gammab}{\max}~\displaystyle\frac{p(\gammab)p(\vec{X}|\gammab)p(\vec{Y}|\vec{X},\gammab)}{p(\vec{Y})}\\&\hspace{-3mm}\overset{(a)}{=}\underset{\vec{X},\gammab}{\max}~\displaystyle{p(\gammab)p(\vec{X}|\gammab)p(\vec{Y}|\vec{X})}\\
&\hspace{-3mm}\overset{}{=}\underset{\vec{X},\gammab}{\min}~\displaystyle{-\log\,p(\vec{Y}|\vec{X})}-\log\,p(\vec{X}|\gammab)-\log\,p(\gammab)\\
&\hspace{-3mm}\overset{(b)}{=}\underset{\vec{X},\gammab}{\min}~\displaystyle\frac{1}{\sigma^2}\|\vec{Y}-\vec{\Phi}\vec{X}\|_{\mathrm{F}}^{2}-\log\,p(\vec{X}|\gammab)-\log\,p(\gammab)\
\end{array}
\end{equation}
where $(a)$ follows from the Markov chain $\gammab\rightarrow\vec{X}\rightarrow\vec{Y}$ and  because $p(\vec{Y})$ does not affect the maximization and $(b)$ follows from the additive Gaussian noise model in \eqref{eq::Y}. The term $p(\vec{X}|\gammab)$  denotes the conditional probability density function (PDF) of the  effective channel $\vec{X}$ given the vector $\gammab$, whereas  the term  $p(\gammab)$ represents the prior belief  on the distribution of the user activity.

Next,  we elaborate in detail the choice of the prior $p(\gammab)$ and the definition of  the conditional PDF $p(\vec{X}|\gammab)$. Then, having fixed these quantities, we derive an  ADMM algorithm to  find an approximate solution to the MAP estimation in  \eqref{eq:map_x_gamma}.

\subsubsection{Sparse prior  $p(\gammab)$}
By the model assumption on the sporadic UE activity, the user activity indicator vector  $\gammab$  exhibits a sparse structure ($\gamma_i=0, \forall i \notin \mathcal{S}$). Thus,  in the context of sparse recovery,  we impose a \emph{sparsity prior $p(\gammab)$} on  $\gammab$. For instance, given a continuous-magnitude random vector $\vecgreek{\theta} \in \mathbb{C}^{N}$, a sparsity-inducing prior can be given by $
p(\vecgreek{\theta})\propto\exp\big( -\textstyle\sum_{i=1}^{N}|\theta_i|^p \big) $, where $p\in [0,1]$ \cite{wipf2004sparse}. 

Note that  setting $p=1$ results in the $\ell_1$-norm penalty corresponding to the \emph{Laplace} density function. On the other hand, setting $p=0$ renders the optimal sparsity-inducing penalty corresponding to the  $\ell_0$-norm. Since $\gammab$ is a  vector of binary elements, setting $p=0$ is equivalent to $p=1$ as it imposes the same sparsity prior $p(\gammab)$. Subsequently, we select  the prior $p(\gammab)$ as  the $\ell_0$-norm penalty as
\begin{equation}\label{eq:pdf_gamma}
p(\gammab) \propto\exp\Big( -\sum_{i=1}^{N}1(\gamma_i)\Big),  
\end{equation}
 
\subsubsection{Conditional probability $p(\vec{X}|\gammab)$}

Since the user activity is controlled by $\gammab$, the conditional probability $p(\vec{X}|\gammab)$ is defined as follows. First,  we note that the  activity patterns of  the different users are mutually independent, hence, the conditional PDF factorizes as $p(\vec{X}|\gammab)=\prod_{i=1}^{N}p(\vec{x}_i|\gamma_i)$. In addition,  for each user $i \in \mathcal{N}$, we distinguish the two possible cases for $p(\vec{x}_{i}|\gamma_i)$ as follows: 1) Conditioned on $\gamma_i=1$, the $i$th UE is active and   $\vec{x}_i$ follows a Gaussian distribution, i.e., $p(\vec{x}_{i}|\gamma_i=1)= p_{\vec{x}_i}$, where $p_{\vec{x}_i}\sim \mathcal{CN}(0,\tilde{\vec{R}}_i)$  and $\tilde{\vec{R}}_i$ denotes the  scaled covariance matrix defined as $\tilde{\vec{R}}_i=p^{\mathrm{UL}}_i\hat{\vec{R}}_i$. 2) Conditioned on $\gamma_i=0$, the $i$th UE is inactive, and $\vec{x}_i$ is a deterministic all-zero vector $\vec{x}_i=\vec{0}$ with probability 1, i.e., $p(\vec{x}_{i}|\gamma_i=0)=1$.
Therefore,  $p(\vec{X}|\gammab)$  is given by
\begin{equation}
p(\vec{X}|\gammab)=\prod_{i=1}^{N}p(\vec{x}_i|\gamma_i)=\prod_{i\in\mathcal{S}}p_{\vec{x}_i}.
\label{eq:pdf_x_cond_gamma}
\end{equation}

By applying the log transformation to  $p(\gammab)$ in \eqref{eq:pdf_gamma} and to $p(\vec{X}|\gammab)$ in \eqref{eq:pdf_x_cond_gamma}, and by dropping the constant terms that do not depend on $\gammab$ and $\vec{X}$, the joint MAP estimation problem \eqref{eq:map_x_gamma} can be equivalently written  as 
\begin{equation}
\label{eq:map_x_gamma_priors}
\{\hat{\vec{X}},\hat{\gammab}\}=\displaystyle\min_{\vec{X},\gammab}\frac{1}{2}\|\vec{Y}-\vec{\Phi}\vec{X}\tran\|_{\mathrm{F}}^{2}+\displaystyle\beta_1\sum_{i=1}^{N}1(\gamma_i )+\beta_2\sum_{i=1}^{N}\vec{x}_{i}\herm\tilde{\vec{R}}_i^{-1}\vec{x}_{i},
\end{equation}
where regularization weights $\beta_1$ and $\beta_2$  balance the emphasis on the priors both in relation to each other and to the measurement fidelity term. The third term in \eqref{eq:map_x_gamma_priors} applies a \textit{quadratic Mahalanobis distance} measure\footnote{The Mahalanobis distance between a vector $\vecgreek{\theta}$ and the Gaussian distribution with mean $\vecgreek{\mu}$ and covariance matrix $\vec{R}$ is defined as $\sqrt{(\vecgreek{\theta}-\vecgreek{\mu})\herm\vec{R}^{-1}(\vecgreek{\theta}-\vecgreek{\mu})}$. It measures the distance between the vector $\vecgreek{\theta}$ and the mean of the distribution ($\vecgreek{\mu}$) measured along the principal component axes determined the covariance matrix $\vec{R}$.}, $ \vec{x}_i\herm\tilde{\vec{R}}^{-1}\vec{x}_i$, $i\in \mathcal{N}$, for active UEs  in order to incorporate the knowledge of the spatial correlation matrices of the UEs into the optimization process. 

\subsection{MAP-ADMM Solution}
The non-convex problem \eqref{eq:map_x_gamma_priors} is a mixed-integer programming problem due to involving binary optimization variables $\gammab$, and is, thus, hard to solve for large $N$. In this section, we  develop a computationally efficient ADMM algorithm, which is numerically illustrated to achieve great performance in Section~\ref{Result}.

We start by noting that the recovery of effective channel $\hat{\vec{X}}$ renders implicitly the vector  $\gammab$, i.e., finding the index set $\{i \mid \gamma_i \neq0,\;i\in\mathcal{N}\}$ is equivalent to finding the index set $\{i \mid \|\vec{x}_i\|_2 >0,\;i\in\mathcal{N}\}$. Therefore, we  solve a relaxed version of the  MAP estimation \eqref{eq:map_x_gamma_priors}  by approximating the penalty term that depend on $\vecgreek{\gamma}$ by  penalty term that depend on $\|\vec{x}_i\|_2,\; \forall i\in \mathcal{N}$.

Note that the second term  $\sum_{i=1}^N 1(\gamma_i)$  in \eqref{eq:map_x_gamma_priors} is equivalent to an $\|\vec{X}\|_{2,0}$ penalty in the sense that it enforces the row-sparsity of the  matrix $\vec{X}\tran$. Therefore,  $\sum_{i=1}^N 1(\gamma_i)$  can be relaxed by the log-sum penalty $\sum_{i=1}^N \log(\|\vec{x}_i\|_2+\epsilon_0)$.
Subsequently, we can eliminate $\gammab$  and approximate  \eqref{eq:map_x_gamma_priors} as
\begin{equation}
      \displaystyle\min_{\vec{X},\vec{u} }\displaystyle\frac{1}{2}\| \vec{Y}-\vec{\Phi} \vec{X}\tran\|_{\mathrm{F}}^2+\beta_1 \sum_{i=1}^N \log(u_i+\epsilon_0)  +\beta_2\sum_{i=1}^{N}\vec{x}_{i}\herm\tilde{\vec{R}}_i^{-1}\vec{x}_{i}\;\;\;
     \mbox{s.t.} \;\;\;\norm{\vec{x}_i} \leq u_i ,~ \forall i \in \mathcal{N}.
   \label{eq::map_approximated}
\end{equation}

Again, we  utilize  MM and  linearize the concave penalty term by its first-order Taylor expansion at point $\vec{u}^{(l)}$. Thus, an approximate solution to \eqref{eq::map_approximated} is found by iteratively solving  the  problem 
\begin{equation}
\begin{array}{ll}\label{eq::map}
\hat{\vec{X}}^{(l+1)}=\displaystyle\min_{\vec{X}}\frac{1}{2}\|\vec{Y}-\vec{\Phi}\vec{X}\tran\|_{\mathrm{F}}^{2}+\beta_1\sum_{i=1}^{N}g_i^{(l)}\norm{\vec{x}_i}+\beta_2\sum_{i=1}^{N}\vec{x}_{i}\herm\tilde{\vec{R}}_i^{-1}\vec{x}_{i},
\end{array}
\end{equation}
where the weight vector $\vec{g}^{(l)}=[g_1^{(l)},g_2^{(l)},\ldots,g_N^{(l)}]\tran$ is given according to \eqref{eq::g}. The optimization problem \eqref{eq::map} can be seen as an iterative reweighted $\ell_{2,1}$-norm minimization  augmented with an additional penalty function that incorporates the spatial correlation  knowledge to the optimization process by applying a Mahalanobis distance penalty on the active UEs.

The objective function in  \eqref{eq::map} is a sum of convex functions, hence, the optimization problem \eqref{eq::map} is convex. Thus, aiming to provide a computationally efficient solution, we develop an ADMM framework that solves  \eqref{eq::map} through a set of sequential update rules, each computed in  closed-form. In particular, in order to decompose \eqref{eq::map} into a set of separate functions,  we introduce two splitting variables $\vec{Z}, \vec{V} \in \mathbb{C}^{M\times N}$  and  rewrite the optimization problem as
\begin{equation}
\begin{array}{ll}
(\hat{\vec{X}}^{(l+1)},\hat{\vec{Z}}^{(l+1)},\hat{\vec{V}}^{(l+1)})=\hspace{-3mm}&\displaystyle\min_{\vec{X},\vec{Z},\vec{V}}\frac{1}{2}\|\vec{Y}-\vec{\Phi}\vec{Z}\tran\|_{\mathrm{F}}^{2}+\beta_1\sum_{i=1}^{N}g_i^{(l)}\norm{\vec{x}_i}+\beta_2\sum_{i=1}^{N}\vec{v}_{i}\herm\tilde{\vec{R}}_i^{-1}\vec{v}_{i}\\ 
      \hspace{-3mm}&\mbox{s.t.} \;\;\;  \quad\vec{x}_i=\vec{z}_i,\;\; \vec{x}_i=\vec{v}_i,~\forall i \in \mathcal{N}.
\end{array}\label{eq::map_blockcvx}
\end{equation}

The optimization problem  \eqref{eq::map_blockcvx}  is \emph{block multi-convex}, i.e., the problem is convex in one set of  variables while  all the other variables are fixed. Since ADMM exploits implicitly the   block multi-convexity nature of \eqref{eq::map_blockcvx},  utilizing ADMM to solve  \eqref{eq::map_blockcvx} is a reasonable choice. Accordingly,
the augmented Lagrangian associated with \eqref{eq::map_blockcvx}  is given by 
\begin{equation}
\begin{array}{ll}
&\mathcal{L}(\vec{X},\vec{Z},\vec{V},\vecgreek{\Lambda}_{\mathrm{z}},\vecgreek{\Lambda}_{\mathrm{v}})=\displaystyle\frac{1}{2}
 \|\vec{Y}- \vec{\Phi} \vec{Z}\tran \|_{\mathrm{F}}^2+\beta_1 \displaystyle\sum_{i=1}^N g_i^{(l)} \norm{\vec{x}_i}+\beta_2 \displaystyle\sum_{i=1}^{N}\vec{v}_i\herm\tilde{\vec{R}}_i^{-1}\vec{v}_i\\&\displaystyle+\frac{\rho}{2}\|\vec{X}-\vec{V}+\frac{1}{\rho}\vecgreek{\Lambda}_{\mathrm{v}}\|_{\mathrm{F}}^2+\displaystyle\frac{\rho}{2}\|\vec{X}-\vec{Z}+\displaystyle\frac{1}{\rho}\vecgreek{\Lambda}_{\mathrm{z}}\|_{\mathrm{F}}^2    -\displaystyle\frac{\fro{\vecgreek{\Lambda}_{\mathrm{z}}}}{2\rho}
     -\displaystyle\frac{\fro{\vecgreek{\Lambda}_{\mathrm{v}}}}{2\rho}.
\end{array}
 \label{eq::Lagrange_MAP}\end{equation}
where $\vecgreek{\Lambda}_{\mathrm{z}}=[\vecgreek{\lambda}_{\mathrm{z}1},\ldots,\vecgreek{\lambda}_{\mathrm{z}N}] \in \mathbb{C}^{M\times N}$ and $\vecgreek{\Lambda}_{\mathrm{v}}=[\vecgreek{\lambda}_{\mathrm{v}1},\ldots,\vecgreek{\lambda}_{\mathrm{v}N}]\in \mathbb{C}^{M\times N}$ are the matrices of the ADMM dual variables.

The ADMM solution to the optimization problem \eqref{eq::map} at the $(l)$th MM iteration is achieved by sequentially minimizing $\mathcal{L}(\vec{X},\vec{Z},\vec{V},\vecgreek{\Lambda}_{\mathrm{z}},\vecgreek{\Lambda}_{\mathrm{v}})$ over the primal variables $(\vec{Z},\vec{V},\vec{X})$, followed by dual variable $(\vecgreek{\Lambda}_{\mathrm{z}},\vecgreek{\Lambda}_{\mathrm{v}})$ updates as follows:
\begin{equation}
\vec{Z}^{(k+1)}:=\displaystyle\min_{\vec{Z}}\mathcal{L}(\vec{X}^{(k)},\vec{Z},\vec{V}^{(k)},\vecgreek{\Lambda}_{\mathrm{z}}^{(k)},\vecgreek{\Lambda}_{\mathrm{v}}^{(k)}):=\displaystyle\min_{\vec{Z}}\frac{1}{2}\| \vec{\Phi}\vec{Z}\tran-\vec{Y}\|_{\mathrm{F}}^2+ \frac{\rho}{2} \Vert\vec{X}^{(k)} -\vec{Z} +\frac{1}{\rho}\vecgreek{\Lambda}_{\mathrm{z}}^{(k)} \|_{\mathrm{F}}^2
\label{eq::z(k+1)_sec2}
\end{equation}
\begin{equation}
\vec{V}^{(k+1)}\!\!:=\displaystyle\min_{\vec{V}}\mathcal{L}(\vec{X}^{(k)},\vec{Z}^{(k+1)},\vec{V},\vecgreek{\Lambda}_{\mathrm{z}}^{(k)},\vecgreek{\Lambda}_{\mathrm{v}}^{(k)})=\!\displaystyle\min_{\vec{V}}\beta_2  \sum_{i=1}^{N}  \vec{v}_i\herm\tilde{\vec{R}}_i^{-1}\vec{v}_i  +\frac{\rho}{2} \Vert   \vec{X}^{(k)} - \vec{V} +\frac{\vecgreek{\Lambda}_{\mathrm{v}}^{(k)}}{\rho} \|_{\mathrm{F}}^2
\label{eq::v(k+1)_sec2}
\end{equation}\vspace{-6mm}
\begin{equation}
\begin{array}{ll}
\!\vec{X}^{(k+1)}\!\!\!\!&:=\displaystyle\min_{\vec{X}}\mathcal{L}(\vec{X},\vec{Z}^{(k+1)},\vec{V}^{(k+1)},\vecgreek{\Lambda}_{\mathrm{z}}^{(k)},\vecgreek{\Lambda}_{\mathrm{v}}^{(k)})\\&:=\displaystyle\min_{\vec{X}}  \sum_{i=1}^{N} \beta_1 g_i^{(l)} \Vert  \vec{x}_i\|_2+\frac{\rho}{2} \|\vec{X}-\vec{Z}^{(k+1)}+ \frac{1}{\rho}\vecgreek{\Lambda}_{\mathrm{z}}^{(k)} \|_{\mathrm{F}}^2 \displaystyle+\frac{\rho}{2} \|\vec{X}-\vec{V}^{(k+1)}+ \frac{1}{\rho}\vecgreek{\Lambda}_{\mathrm{v}}^{(k)} \|_{\mathrm{F}}^2
\end{array}
\label{eq::x(k+1)_sec}
\end{equation}\vspace{-3mm}
\begin{equation}
\hspace{-2.5cm}\vecgreek{\Lambda}_{\mathrm{z}}^{(k+1)} := \vecgreek{\Lambda}_{\mathrm{z}}^{(k)}+\rho\big(\vec{X}^{(k+1)}  -\vec{Z}^{(k+1)} \big),
\label{eq::lambda_z(k+1)}
\end{equation}\vspace{-8mm} 
\begin{equation}
\hspace{-2.5cm}\vecgreek{\Lambda}_{\mathrm{v}}^{(k+1)} := \vecgreek{\Lambda}_{\mathrm{v}}^{(k)}+\rho\big(\vec{X}^{(k+1)}  -\vec{V}^{(k+1)} \big),
\label{eq::lambda_v(k+1)}
\end{equation} 

We present the derivations of the ADMM steps \eqref{eq::z(k+1)_sec2}, \eqref{eq::v(k+1)_sec2}, and \eqref{eq::x(k+1)_sec} in detail below. 
\paragraph{$\vec{Z}$-update} We note that the  $\vec{Z}$-update in \eqref{eq::z(k+1)_sec2}  is identical to the convex optimization problem in \eqref{eq::z(k+1)}, hence, $\vec{Z}^{(k+1)}$ is computed using \eqref{eq::Z+}.

\paragraph{$\vec{V}$-update}  We can easily show that the $\vec{V}$-update  in \eqref{eq::v(k+1)_sec2} can be decoupled into $N$ convex sub-problems, given by
\vspace{-.2cm}
\begin{equation}
   \vec{v}_i^{(k+1)}= \min_{\vec{v}_i}\beta_2\vec{v}_i\herm\tilde{\vec{R}}_i^{-1}\vec{v}_i+\displaystyle\frac{\rho}{2}\big \|\vec{x}^{(k)}_i-\vec{v}_i+\displaystyle\frac{\vecgreek{\lambda}_{\mathrm{v}i}^{(k)}}{\rho}\big\|_2^2,\;\; \forall i \in  \mathcal{N}. \label{eq::v_map}
\end{equation}\vspace{-.2cm}
The solution for \eqref{eq::v_map} is obtained by setting the derivative of the objective function with respect to $\vec{v}_i$ to zero, resulting in 
\begin{equation}
\vspace{-.2cm}
     \vec{v}_i^{(k+1)}=  \frac{1}{\beta_2}\tilde{\vec{R}}_i\big(\frac{\rho}{\beta_2}\tilde{\vec{R}}_i+\vec{I}_M \big)^{-1}(\rho \vec{x}_i^{(k)}+\vecgreek{\lambda}_{\mathrm{v}i}^{(k)}), ~\forall i \in \mathcal{N}
    \label{eq::v+}
\end{equation}

\paragraph{$\vec{X}$-update} Using the manipulations for \eqref{eq::x(k+1)_sec} shown in Appendix A, the  $\vec{X}$-update  solves the following convex optimization problem:
\vspace{-2mm}
\begin{equation}
    \vec{X}^{(k+1)}:=\min _{\vec{X}} \sum_{i=1}^{N} \alpha_i^{(k)} \norm{\vec{x}_i} +\rho  \|   \vec{X}-  \vec{S}^{(k)}\| _{\mathrm{F}}^2,  
  \label{eq::X++}
\end{equation}
where  $\alpha_i^{(k)}= \beta_1 g_i^{(l)}$ and $\vec{S}^{(k)}=\dfrac{1}{2}\big( \vec{Z}^{(k+1)}+\vec{V}^{(k+1)}-\displaystyle\frac{\vecgreek{\Lambda}_{\mathrm{z}}^{(k)}+\vecgreek{\Lambda}_{\mathrm{v}}^{(k)}}{\rho}\big)$. The problem \eqref{eq::X++} decouples into  $N$ sub-problems, each admitting the closed-form solution 
\begin{equation}
\vec{x}_i^{(k+1)}= \frac{\max{\big\{0,\norm{\vec{s}_i^{(k)}}-\frac{\alpha_i^{(k)}}{2\rho}\big\}}}{\norm{\vec{s}_i^{(k)}}}\vec{s}_i^{(k)},\quad \forall i \in \mathcal{N}.
    \label{eq::prox2}
\end{equation}

\subsection{CDI Knowledge}\label{sec:CDI}
MAP-ADMM operates on the assumption that the BS knows the  CDI of the individual channels, i.e., $\{\vec{R}\}_{i=1}^N$. We note that the assumption that the BS knows $\{\vec{R}_i\}_{i=1}^N$  is widely accepted in the massive MIMO literature  \cite{massivemimobook,Li-etal15}. Furthermore, a similar assumption on the availability of the CDI  has been adopted  in \cite{cheng2020orthogonal} that addresses JUICE in mMTC with sporadic user activity.

The acquisition of the CDI at the BS may be challenging,  especially for the UEs which are inactive for a long period. Therefore, a possible solution to  circumvent such an issue can be realized by deploying a training phase to estimate the CDI. The training phase can be implemented over separate channel resource blocks that are solely dedicated to estimate the CDI. In practice, the BS would consume a set of available channel resources in order to obtain an estimate of all the channel covariance matrices, denoted as $\{\hat{\vec{R}}_i\}_{i=1}^N$. In particular, at different time intervals, a specific group  of UEs  transmit pre-assigned orthogonal training  pilots to the BS over $T$ coherence intervals, and subsequently, the BS employs conventional MIMO channel estimation techniques to obtain $T$ estimates of channel responses $\vec{h}_i$, denoted as  $\hat{\vec{h}}_i^{1},\ldots,\hat{\vec{h}}_i^{T}$. Subsequently, the BS computes the estimated channel covariance matrix\footnote{
For a recent review on the different techniques on channel estimation and   channel covariance estimation in massive MIMO networks,  we refer the reader to \cite[Sect.~3.2-3.3]{massivemimobook}.} for the $i$th UE as  $\hat{\vec{R}}_i=\frac{1}{T}\sum_{t=1}^{T}\hat{\vec{h}}_i^{t}\hat{\vec{h}}_i^{t\herm}$.

The frequency of updating the CDI at the BS depends on the mobility and the activity level of the UEs as well as the changes in the multi-path environment. Therefore, the estimated $\hat{\vec{R}}_i$ can be used over several coherence intervals  due to the fact that: 1)  the channel covariance matrices vary in a slower timescale compared to the channel coherence time, and 2) the  UEs have  very low mobility  in many practical mMTC systems. Consequently, learning the  CDI does not consume disproportionate amount of resources. As we will show  in the simulation results, the BS does not require a large number of training samples $T$ to estimate $\{\vec{R}\}_{i=1}^N$.  In fact, the BS needs roughly $T=2M$ samples to provide near-optimal results in terms of the  mean square error for channel estimation, and we note that similar conclusion has been reported in \cite[Sect.~3.3.3]{massivemimobook}.  
\subsection{Algorithm Implementation}
The details of the proposed MAP-based JUICE, termed MAP-ADMM, are summarized in Algorithm 2. We note that $\vec{Z}$-update \eqref{eq::z(k+1)_sec2} and the $\vec{V}$-update \eqref{eq::v(k+1)_sec2} are independent from each other, hence, they can be performed fully in parallel. Similarly to Algorithm 1, MAP-ADMM is run until $\fro{\vec{X}^{(k)}-\vec{X}^{(k-1)}}<\epsilon$  or until a maximum number of iterations  $l_{\mathrm{max}}k_{\mathrm{max}}$ is reached. 

\begin{algorithm}[t]
\DontPrintSemicolon
   \KwInput{1) Pilot matrix $\vec{\Phi}$, covariance matrices $\{\tilde{\vec{R}}_i\}_{i=1}^N$, $\big( \vec{\Phi}\tran\vec{\Phi}^*+\rho \vec{I}_N\big)^{-1}$, $\frac{1}{\beta_2}\tilde{\vec{R}}_i\big(\frac{\rho}{\beta_2}\tilde{\vec{R}}_i+\vec{I}_M \big)^{-1}, \forall i \in \mathcal{N}$ \\  2) parameters $\beta_1$,$\beta_2$ ,$\rho$, $\epsilon_0$, $\epsilon$, $\kappa$, $l_\mathrm{max}$,$k_\mathrm{max}$, }
  \KwOutput{$\hat{\vec{X}}$}
 \Kwinitialize{$\vec{X}^{(0)},\vec{V}^{(0)},\vec{Z}^{(0)},\vecgreek{\Lambda}_{\mathrm{v}}^{(0)},\vecgreek{\Lambda}_{\mathrm{z}}^{(0)}, k=1, l=1.$}
 Receive $\vec{Y}$ at the BS, and compute and store  
   \While{$l<l_{\mathrm{max}}$  }
   {
   \While{$k<k_{\mathrm{max}}$ $\mathrm{or}$ $\| \vec{X}^{(k)}-\vec{X}^{(k-1)} \|<\epsilon$}
   {
Update $\vec{Z}^{(k+1)}$ using  \eqref{eq::Z+} and $\vec{V}^{(k+1)}$ using \eqref{eq::v+}\;
Update $\vec{X}^{(k+1)}$ using  \eqref{eq::prox2} 
\;
 $\vecgreek{\Lambda}_{\mathrm{z}}^{(k+1)}=  \vecgreek{\Lambda}_{\mathrm{z}}^{(k)}+\rho\big(   \vec{X}^{(k+1)}-\vec{Z}^{(k+1)} \big)$\;
 $\vecgreek{\Lambda}_{\mathrm{v}}^{(k+1)}=  \vecgreek{\Lambda}_{\mathrm{v}}^{(k)}+\rho\big(   \vec{X}^{(k+1)}-\vec{V}^{(k+1)} \big)$\;
$k\leftarrow{k+1}$\;
}
   $\vec{X}^{(l)}\leftarrow \vec{X}^{(k+1)}$\;
  Update  $ g_{i}^{(l)}$ using \eqref{eq::g} \;
$l\leftarrow{l+1}$\;
}
\caption{MAP-ADMM}
\end{algorithm}

\section{Algorithm Computational Complexity}
In a typical mMTC scenario, where the number of connected devices is very large, the complexity of the recovery algorithms is an important issue to address. In fact, for the implementation of the proposed algorithms,  the computational complexity determines the hardware processing cost. 
Next, we analyze the complexity of the proposed JUICE algorithms in terms of the number of required complex multiplications per iteration using the big $\mathcal{O}(\cdot)$ notation. The complexity analysis is summarized in Table \ref{table:1} which also shows the exact number of matrix multiplications.

At the $\vec{Z}$-update step of IRW-ADMM and MAP-ADMM, for fixed $\rho$, the quantity $\big( \vec{\Phi}\tran \vec{\Phi}^*+\rho \vec{I}_N\big)^{-1}$ is computed only once at an algorithm initialization. Similarly, the term $\vec{Y}\tran\vec{\Phi}^*$ is computed only once  upon receiving the pilot signal $\vec{Y}$. Therefore, computing $\vec{Z}^{(k+1)}$ requires $(M+1)N^2$ complex multiplications. For the $\vec{V}$-update of MAP-ADMM, the terms $\frac{1}{\beta_2}\tilde{\vec{R}}_i\big(\frac{\rho}{\beta_2}\tilde{\vec{R}}_i+\vec{I}_M \big)^{-1}, \forall i \in \mathcal{N}$, in \eqref{eq::v+} need to be computed only once and can subsequently be used for several coherence intervals. Hence, MAP-ADMM requires $N(M^2+2M)$ complex multiplications to compute $\vec{V}^{(k+1)}$.
The soft-threshold operators in \eqref{eq::prox} and \eqref{eq::prox2} for the $\vec{X}$-update require $2MN$ complex multiplications. Finally, the weight vector $\vec{g}^{(l)}$ is computed only at the outer MM iteration level $(l)$ and it requires  $MN$ complex multiplications. Therefore, the overall complexity for IRW-ADMM and MAP-ADMM is $\mathcal{O}(MN^2)$ and $\mathcal{O}(MN^2+NM^2)$, respectively. 

Table \ref{table:1} also compares the complexity of IRW-ADMM and MAP-ADMM to the three baseline algorithms that we consider in the numerical experiments: Fast alternating direction method (F-ADM) \cite{lu2011fast}, simultaneous orthogonal matching pursuit (SOMP) \cite{tropp2006algorithms}, and temporal sparse Bayesian learning (T-SBL) \cite{zhang2011sparse}. F-ADM  solves the problem  \eqref{eq::l2,1} using an ADMM algorithm and it has computational complexity of $\mathcal{O}(M\tau_{\mathrm{p}}N)$. The greedy SOMP is reported  in \cite{ke2020compressive} to exhibit computational complexity of $\mathcal{O}(M\tau_{\mathrm{p}}N)$. T-SBL 
has computational complexity\footnote{The authors in \cite{zhang2011sparse} also devised a low-complexity version of T-SBL relying on approximate updates which was shown to work well in the high SNR regime. However, since we are interested in a broader SNR range, this implementation is not readily applicable to our JUICE problem.} of $\mathcal{O}(N^2M^3\tau_{\mathrm{p}})$.

In summary, incorporating the channel spatial correlation information results in increased computational complexity. For instance, MAP-ADMM has higher computational complexity per iteration compared to IRW-ADMM due to incorporating the spatial structure information in  the $\vec{V}$-update step. In addition, as the  proposed IRW-ADMM and MAP-ADMM aim at providing an exact solution to the JUICE problem, they are more computationally complex than F-ADM. Nevertheless, the additional cost of the proposed algorithms is compensated for by the convergence to a more accurate solution, as we will show in the next section.

\begin{table}[h!]

\centering
\caption{Computational complexity for different recovery algorithms, where $(k)$ is the iteration index and $\hat{K}$ is the estimated number of non-zero element at any particular iteration}
\begin{tabular}{c|| c||c } 
 \hline
 Algorithm &  Number of complex multiplications in each iteration & $\mathcal{O}(\cdot)$ \\ [0.5ex] 
 \hline \hline
IRW-ADMM & $(M+1)N^2+3M\hat{K}+M(N-\hat{K})+MN$& $\mathcal{O}(MN^2)$  \\ 
MAP-ADMM & $(M+1)N^2+NM^2+3M\hat{K}+M(N-\hat{K})+MN$& $\mathcal{O}(MN^2+NM^2)$  \\   
F-ADM \cite{lu2011fast} &  $4\tau_{\mathrm{p}}MN+5MN$ &$\mathcal{O}(\tau_{\mathrm{p}}MN)$ \\
SOMP \cite{tropp2006algorithms} & $(2\tau_{\mathrm{p}}+1)MN+\tau_{\mathrm{p}}M^2+(M+1)\tau_{\mathrm{p}}(k)^2+(k)^3$  &$\mathcal{O}(\tau_{\mathrm{p}}MN)$ \\
T-SBL \cite{zhang2011sparse} & $2M^3N^2\tau_{\mathrm{p}}+M^2\tau_{\mathrm{p}}^2+NM^2\tau_{\mathrm{p}}$&$\mathcal{O}(N^2M^3\tau_{\mathrm{p}})$\\
 \hline
\end{tabular}
\label{table:1}
\end{table}

\section{Simulation Results}
\label{Result}
In this section, we provide simulation results to show the performance of  the proposed JUICE algorithms in terms of user activity detection accuracy,  channel estimation quality, and convergence rate, and compare them to existing MMV reconstruction algorithms. 

\subsection{Simulation Setup}
We consider a single-cell of a radius of $50 $~m, where the BS is surrounded by $N=200$ uniformly distributed UEs, out of which $K=10$ UEs are active at each coherence interval  $T_\mathrm{c}$.  The propagation channel between the $i$th user and the BS in \eqref{eq::ch} consists of $P_i=200$ paths with 
angular spread deviation $\sigma_{\psi}= 10^{\circ}$. Each user $i=1,\ldots,N$ is assigned with a unique normalized quadratic phase shift keying (QPSK) sequence $\vecgreek{\phi}_i$, where the  QPSK pilot symbols are drawn from  an i.i.d. complex Bernoulli distribution.  The SNR is defined as  
 $\text{SNR}\, \left[\text{dB} \right]=10 \log_{10}\left(\frac{\mathbb{E}[\Vert \vec{\Phi}\vec{X}\Vert_{\mathrm{F}}^{2}]}{\mathbb{E}\left[\Vert \vec{W}\Vert_{\mathrm{F}}^{2} \right]}\right).$
\subsection{Performance Metrics}
The JUICE performance  is quantified in terms of normalized mean square error (NMSE),  support recovery rate (SRR), and the convergence rate. The  NMSE is defined  as $\frac{\mathbb{E}\left [\Vert \vec{X}_\mathcal{S}-\hat{\vec{X}}_\mathcal{S}\Vert_{\mathrm{F}}^2 \right ]}{\mathbb{E}\left[\Vert \vec{X}_\mathcal{S}\Vert_{\mathrm{F}}^2 \right]}$, 
where $\vec{X}_\mathcal{S}$ and $\hat{\vec{X}}_\mathcal{S}$ denote the original and estimated effective channel matrix, respectively, restricted to the true active support $\mathcal{S}$. 
The expectation in the NMSE is computed via Monte-Carlo averaging over the randomness of effective channel matrix $\vec{X}$, the pilot sequence matrix  $\vec{\Phi}$, and noise $\vec{W}$; thus, the NMSE is presented as the \textit{normalized average square error} (NASE).

The  SRR is defined as $
    \frac{\vert \mathcal{S} \cap \hat{\mathcal{S}}\vert}{\vert \mathcal{S} - \hat{\mathcal{S}}\vert+K}$,
where $\hat{\mathcal{S}}=\{i\,\mid \;\Vert \hat{\vec{x}}_i\Vert_2>\epsilon_{\mathrm{thr}},\; \forall i \in \mathcal{N}\}$ denotes the detected support for a small pre-defined threshold $\epsilon_{\mathrm{thr}}$. Thus, $\vert \mathcal{S} \cap \hat{\mathcal{S}}\vert$ represents the number of correctly identified active users, whereas $\vert \mathcal{S} - \hat{\mathcal{S}}\vert$  accounts for both the number of misdetected active UEs and falsely identified inactive  UEs. The SRR rate approaches 1 when $\hat{\mathcal{S}}$ is  close to the true   $\mathcal{S}$. 

\subsection{Baselines}\label{sec_results_baselines}
We compare  the performance of the proposed algorithms against the following algorithms that solve any  MMV sparse recovery problem: 
1) SOMP \cite{tropp2006algorithms}, 
2) F-ADM algorithm \cite{lu2011fast}, which differs from the proposed ADMM in Algorithm 1 (with $\vec{g}^{(l)}=\vec{1}$) in that while ADMM provides an exact solution to \eqref{eq::l2,1}, F-ADM solves  \eqref{eq::l2,1} approximately by linearizing the $\vec{X}$-update sub-problem with the first-order Taylor expansion; 
3) SPARROW, which reformulates \eqref{eq::l2,1} as a semi-definite programming problem  \cite[Eq.~(22)]{steffens2018compact} and we solve it using  CVX toolbox \cite{cvx}; and 
4) T-SBL\cite{zhang2011sparse} where both the second-order statistics and the noise variance are known at the BS and the sparse recovery is performed using the update rules given by \cite[Eqs.~(6),~(7),~(12)]{zhang2011sparse} (i.e., ``$\vec{B}$-update'' in \cite[Eq.~(13)]{zhang2011sparse} is not performed, because we provide the covariance matrices  $\{\vec{\hat{R}}_i\}_{i=1}^N$.). In addition, we use both the oracle least square (LS) and the  oracle joint minimum mean square error (MMSE) estimator, shown in Appendix B, where each estimator is provided ``oracle'' knowledge on the true set of active UEs. While the oracle LS estimator provides a good benchmark for channel estimation when no CDI is available at the BS, the joint MMSE estimator provides a lower bound on  channel estimation performance when both the CDI and the noise variance is available at the BS.

\vspace{-.4cm}
\subsection{Parameter Tuning}
The sparse recovery algorithms require fine-tuning of their regularization parameters to yield their best estimates. While the regularization parameters depend on the different system parameters, such as $N$, $M$, $K$, $\tau_{\mathrm{p}}$, and $\sigma^2$, they are selected empirically by cross-validation in practice.  For a fair comparison, all the parameters have been empirically tuned in advance and then fixed such that they provide overall the best performance in terms of NASE for the SNR range $[0-16]~$dB. For instance, $\beta_1$ depends highly on the ratio $\frac{K}{N}$, however, since the $K$ is not known to the BS in general, we tuned  $\beta_1$ based on the noise variance $\sigma^2$ as $\beta_1=\sqrt{\frac{\sigma^2}{2}}$ since it provided the most robust convergence.  Furthermore, we set  $\beta_2$ and log-sum stability parameter $\epsilon_0$ to $\beta_2=1~\%$ and $\epsilon_0=0.07-0.12~\%$  of the average norm of the effective channels.  Moreover, since ADMM converges typically in few tens of iterations, a maximum number of iterations of  $l_{\mathrm{max}}=12$, $k_{\mathrm{max}}=5$, and stopping criterion  $\epsilon=10^{-3}$ were found to be sufficient for the ADMM-based algorithms to converge to their best performance. All the optimization variables for MAP-ADMM ($\vec{X}, \vec{V},\vec{Z}, \vec{\Lambda}_{\mathrm{z}}$, and $\vec{\Lambda}_{\mathrm{v}}$) and for IRW-ADMM ($\vec{X}$, $\vec{Z}$, and $\vec{\Lambda}$) are initialized  as zero matrices. The results are obtained by averaging over $10^3$ random channel realizations.  

\vspace{-.4cm}
\subsection{Results}
\subsubsection{Performance without Side Information}
First, we examine the scenario when no CDI is available to the BS. To this end,  we compare  the performance of the proposed ADMM and  IRW-ADMM algorithms (Algorithm 1) with F-ADM, SPARROW, SOMP, and oracle LS. 

\begin{figure}[t]
     \centering
     \begin{subfigure}{0.49\textwidth}
    \includegraphics[scale=.5]{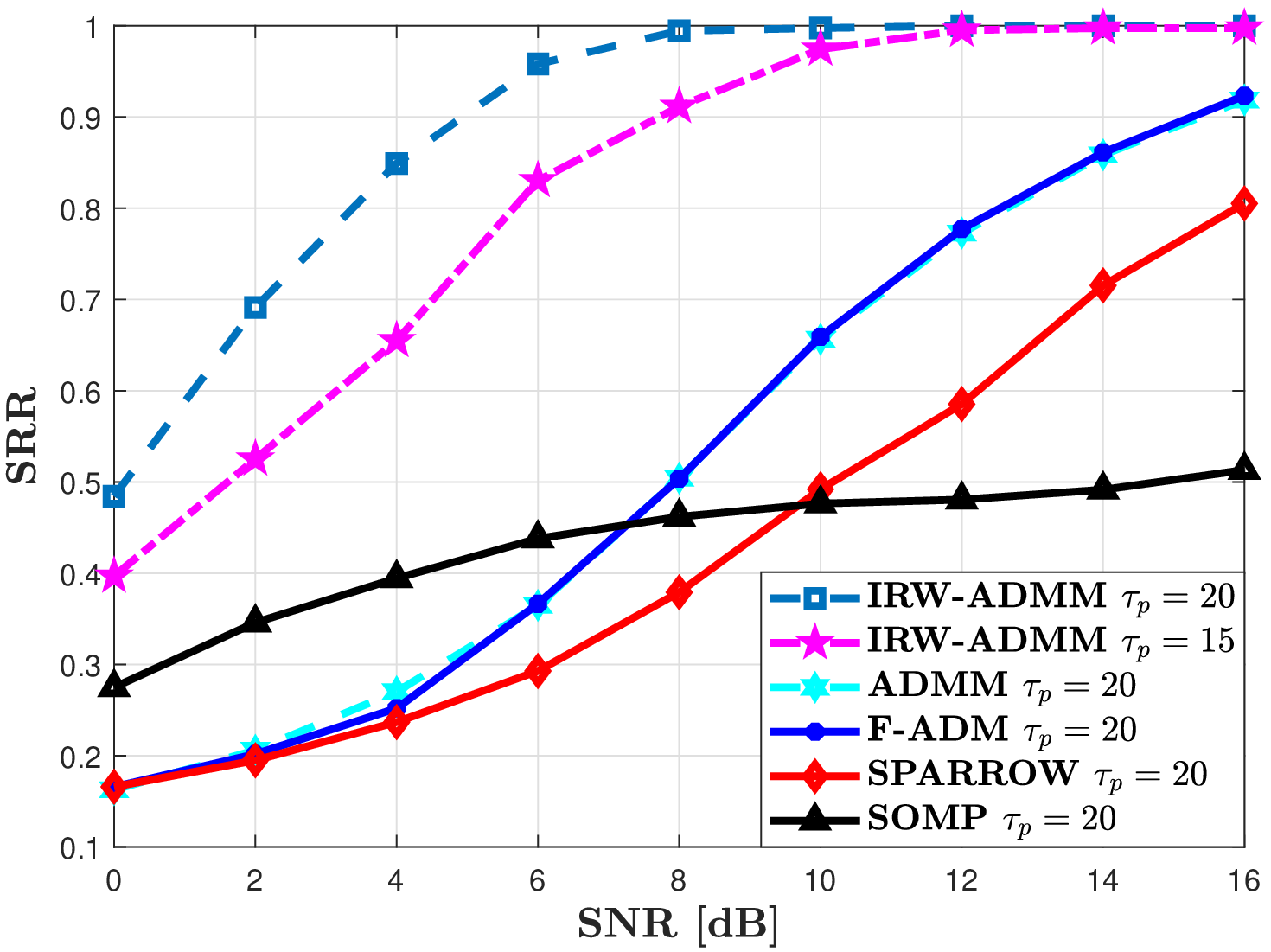}
    \caption{ SRR versus SNR.}
    \label{fig:SRR_P1}
\end{subfigure}
   \hfill      
   \begin{subfigure}{0.49\textwidth}
  \centering
     \includegraphics[scale=.5]{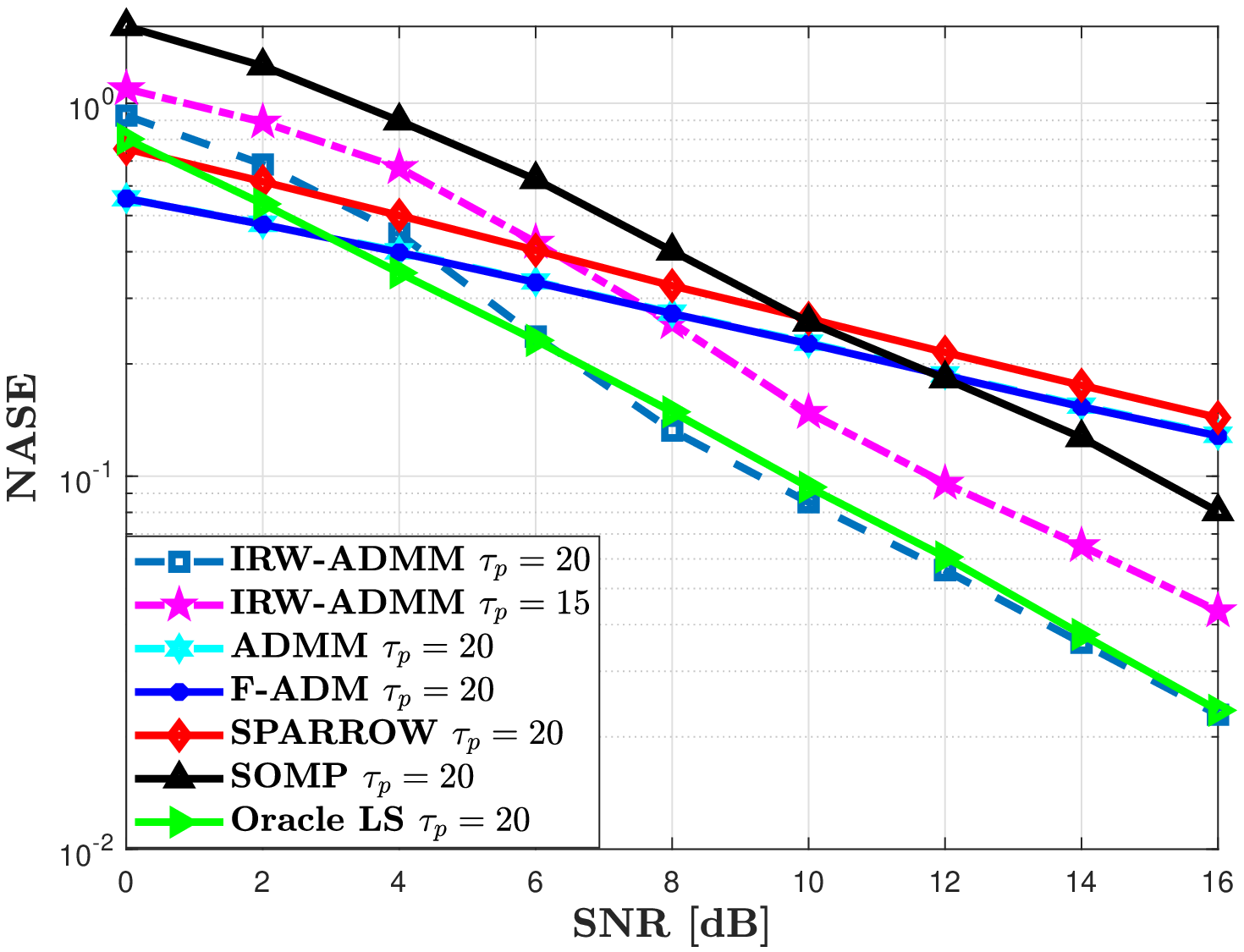}
     \caption{NASE rate SNR.}
     \label{fig:NASE_P1}
\end{subfigure}
\caption{ Performance of JUICE with no side information  in terms of SRR and NASE against  SNR for $N=200$, $M=20$, and $K=10$.}
\label{fig:mmv_perfromance}\vspace{-6mm}
\end{figure}

Fig.\ \ref{fig:mmv_perfromance}(a) illustrates the obtained SRR against SNR for the different sparse recovery algorithms. The obtained results reveal that the proposed IRW-ADMM provides the best performance by achieving the highest user activity detection accuracy. In fact, the IRW-ADMM using  pilot sequence length $\tau_{\mathrm{p}}=20$  is able to achieve SRR $=0.95$ for ${\text{SNR}>6}$~dB. Furthermore, even with the $25~\%$ reduced pilot length, i.e., $\tau_{\mathrm{p}}=15$, the IRW-ADMM still outperforms the other MMV recovery algorithms by a large margin.  Fig.\ \ref{fig:mmv_perfromance}(a) shows that the proposed ADMM provides similar performance F-ADM. However, we note that ADMM  uses fewer regularization parameters compared to F-ADM, thus, it may be  more resilient to parameter tuning. 

Fig.\ \ref{fig:mmv_perfromance}(b) depicts the channel estimation performance  for the different recovery algorithms  in terms of NASE  against SNR, including the comparison  to the genie-aided LS benchmark.  It can be readily seen that for $\tau_{\mathrm{p}}=20$,   the performance of the proposed IRW-ADMM nearly matches the performance of the genie-aided LS. Furthermore,   IRW-ADMM with a reduced  pilot sequence length of $\tau_{\mathrm{p}}=15$ still outperforms ADMM, F-ADM, SOMP and SPARROW for  SNR $>8$~dB. Similarly to the SRR performance, the proposed ADMM and F-ADM achieve similar NASE performance.  Moreover, as the sparsity regularization parameter for both F-ADM and ADMM  is based on the knowledge of the noise variance, F-ADM and ADMM shows to outperform the oracle LS for SNR $<4$~dB. Finally,  while SOMP provides a lower SRR performance, it outperforms ADMM, F-ADM, and SPARROW in terms of NASE for the high SNR regime. The low  SRR  in SOMP is caused by the high number of falsely identified inactive  UEs. However, since NASE is quantified only for the true active UEs, the NASE performance does not suffer a huge degradation. In summary, the results presented in Fig.\ \ref{fig:mmv_perfromance}  highlight the remarkable gains obtained by formulating the JUICE as an  iterative reweighted $\ell_{2,1}$-norm minimization problem.

\begin{figure}[t]
     \centering
     \includegraphics[scale=.55]{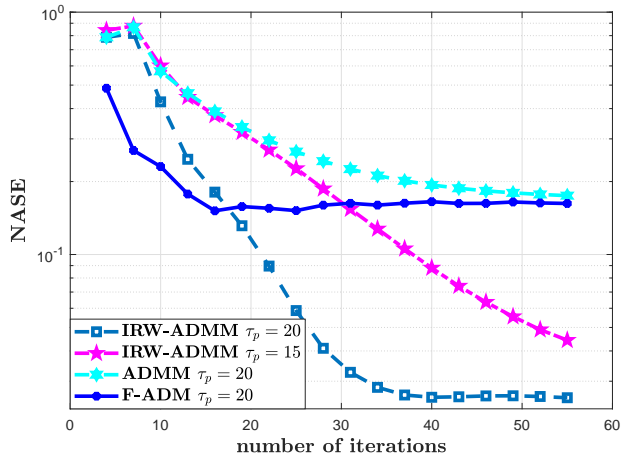}\vspace{-3mm}
     \caption{Convergence behaviour in terms of NASE versus the number of iterations at SNR $=16$~dB for $N=200$, $M=20$, and $K=10$.}
     \label{fig:iter_mmv}\vspace{-5mm}
 \end{figure}

Fig.\ \ref{fig:iter_mmv}  presents a typical convergence behavior of  ADMM, IRW-ADMM, and F-ADM for $\text{SNR}=16$ dB. The figure shows  the number of iterations required for the algorithms to converge to the optimal performance. The results reveal that  IRW-ADMM using $\tau_\mathrm{p}=20$ takes approximately $40$ iterations to convergence, with a slower convergence when reducing the pilot sequence length  to $\tau_\mathrm{p}=15$. On the other hand, different from the SRR and NASE performance where ADMM and F-ADM provide similar performance,  F-ADM  converges in about 20 iterations which is much faster than the proposed ADMM taking about $50$ iterations to converge.

\subsubsection{ The Impact of Exploiting Channel Statistics}
Since we have shown the superiority of IRW-ADMM over conventional sparse recovery algorithms where no knowledge on the CDI is used, next, we  investigate the effect of incorporating the  CDI on the JUICE performance. 

\begin{figure}[t]
     \centering
     \begin{subfigure}{0.49\textwidth}
    \includegraphics[scale=.5]{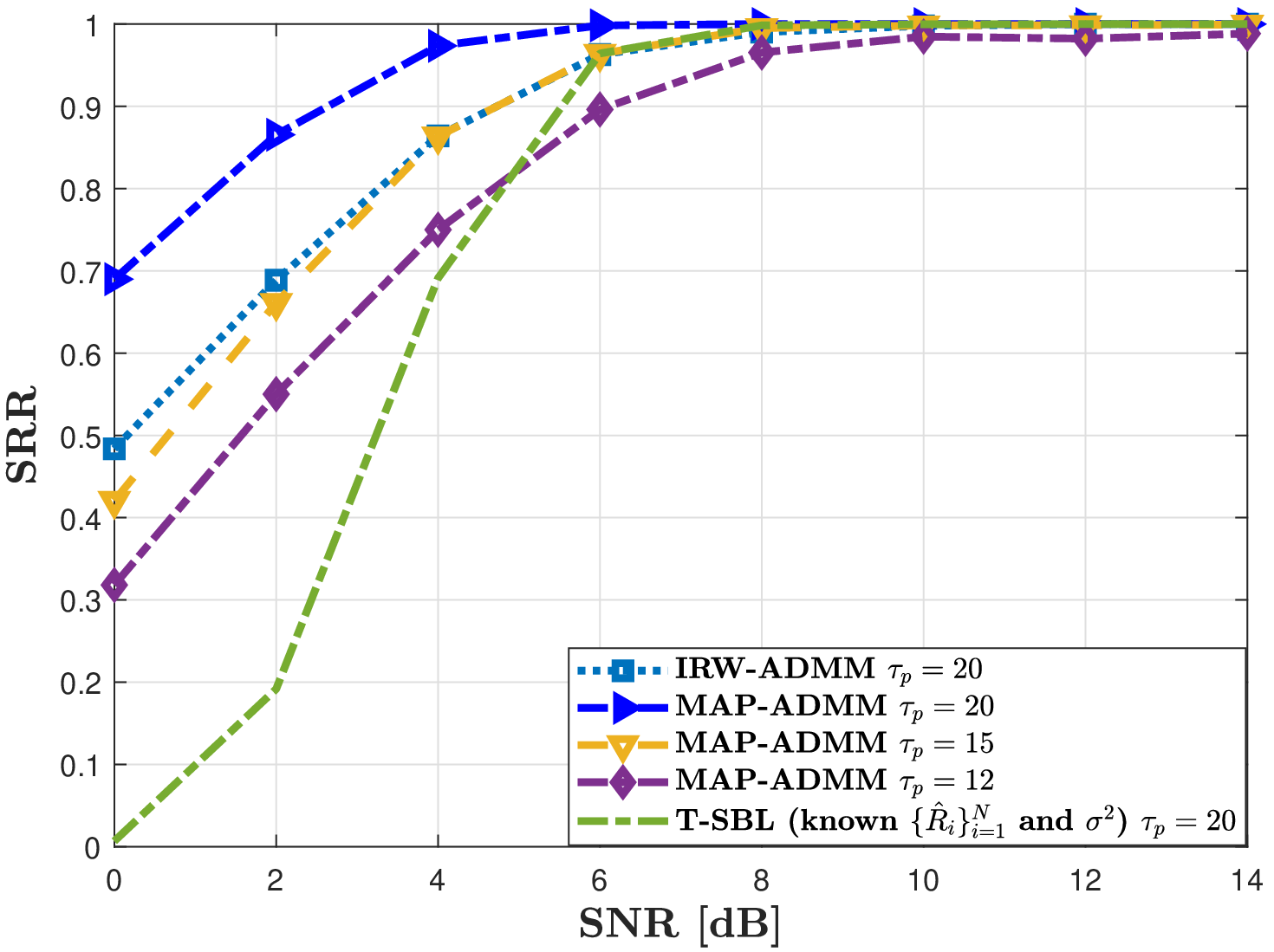}
    \caption{ SRR versus SNR.}
    \label{fig:SRR_P1}
\end{subfigure}
   \hfill      
   \begin{subfigure}{0.49\textwidth}
  \centering
     \includegraphics[scale=.5]{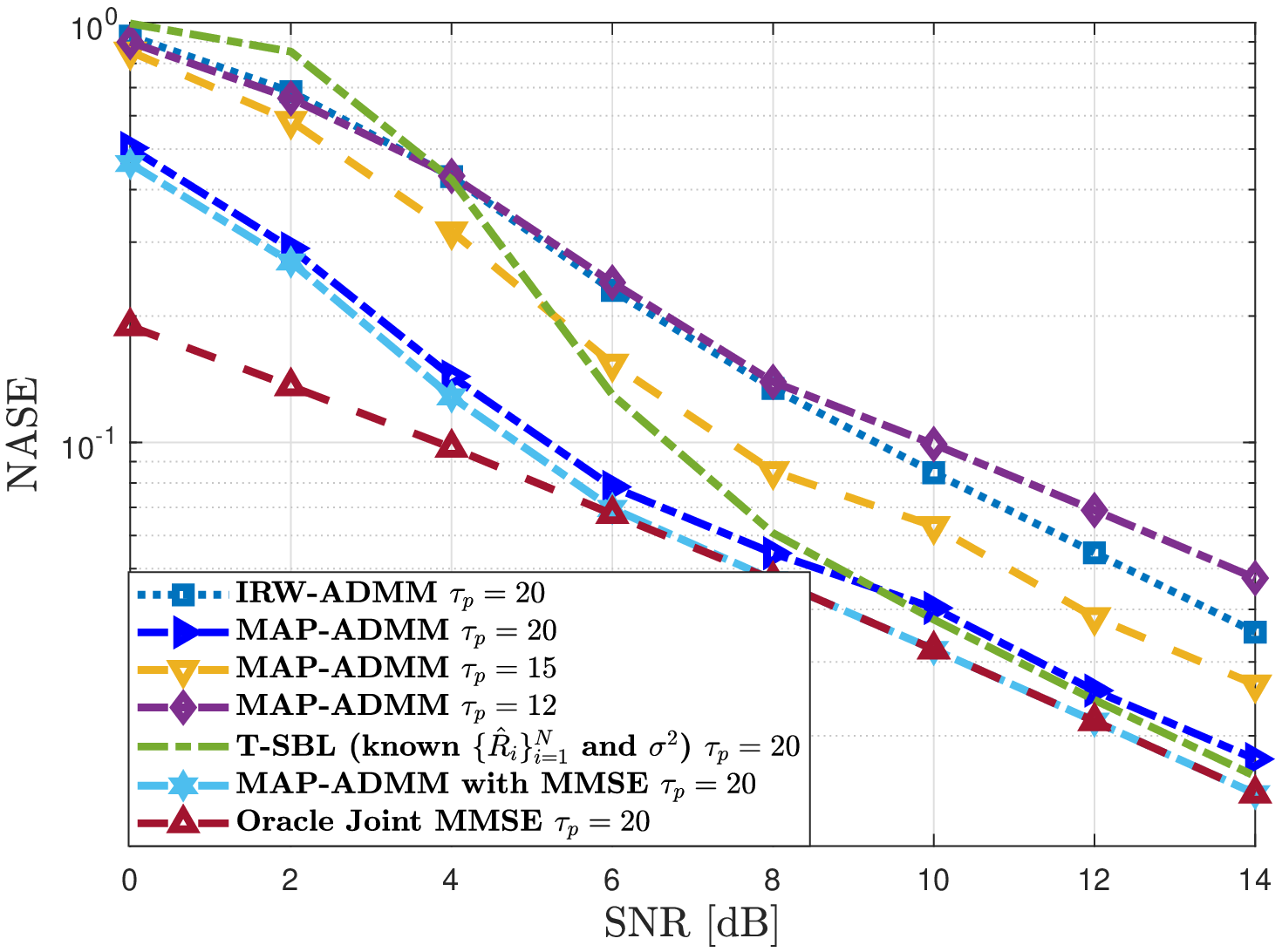}
     \caption{NASE rate SNR.}
     \label{fig:NASE_P1}
\end{subfigure}
\caption{ Performance of the proposed MAP-ADMM   in terms of SRR and NASE against  SNR for $N=200$, $M=20$, and $K=10$.}
\label{fig:mmv+CDI_perfromance}\vspace{-6mm}
\end{figure}

First, we quantify the activity detection accuracy performance of the proposed MAP-ADMM and compare it to IRW-ADMM and T-SBL. Fig.\ \ref{fig:mmv+CDI_perfromance}(a) presents the SRR  against SNR for the proposed algorithms for different values of pilots lengths. The results clearly show that  MAP-ADMM provides superior performance compared to IRW-ADMM. For instance, MAP-ADMM  identifies the set   of true active users $\mathcal{S}$ perfectly for  SNR $>8$~dB using a pilot length $\tau_{\mathrm{p}}=20$. Furthermore, reducing the pilot length by a factor of $25~\%$ (i.e., $\tau_{\mathrm{p}}=15$) affects the performance of MAP-ADMM only  moderately and optimal performance is achieved for   SNR $>10$~dB. More interestingly,  the results indicate
that even with $40~\%$  reduction in the pilot sequence length (i.e., $\tau_{\mathrm{p}}=12$),  MAP-ADMM provides  $95~\%$ SRR rate for SNR $>10$~dB. Finally, the results show that while T-SBL suffers from an inferior performance in the low SNR regime, it provides an optimal activity detection performance when SNR $>8$~dB.

Fig.\ \ref{fig:mmv+CDI_perfromance}(b) illustrates the channel estimation performance in terms of  NASE   for MAP-ADMM  against SNR for different pilot lengths and compares it to IRW-ADM, T-SBL,  and the oracle MMSE  benchmark. The proposed MAP-ADMM indisputably provides superior performance and significant improvement over  IRW-ADMM. For instance, given the same pilot sequence length of $\tau_{\mathrm{p}}=20$, MAP-ADMM achieves the same performance as  IRW-ADMM while using up to  $6$ dB lower SNR. Furthermore, Fig.\ \ref{fig:mmv+CDI_perfromance}(b) reveals  one advantageous feature of utilizing available CDI: even with  25 $\%$ reduction in the pilot length, i.e., $\tau_{\mathrm{p}}=15$, MAP-ADMM still provides 2 dB gain compared to  IRW-ADMM.  Comparing the performances between the MAP-ADMM and T-SBL algorithms, we distinguish  two cases: 1) For ${\text{SNR}<8}$~dB, T-SBL does not provide a reliable performance and MAP-ADMM outperforms T-SBL by a large margin, or so. 2) In  the high SNR regimes, i.e., $\text{SNR} \geq 8$~dB, T-SBL outperforms slightly MAP-ADMM. These results  can be explained by the fact that, in contrast to MAP-ADMM, T-SBL knows and uses the exact noise variance $\sigma^2$. However, when the BS has the exact  knowledge on the noise variance $\sigma^2$ as well as the CDI, the slight gap in the NASE performance between T-SBL and MAP-ADMM  can be compensated for by utilizing a joint MMSE estimator on the received signal associated with the estimated active UE set $\hat{\mathcal{S}}$ obtained by  MAP-ADMM. Fig.\ 4(b)  shows that using the joint MMSE estimator on the  estimated active UEs  provides even the same performance as the  oracle joint MMSE estimator starting from  ${\text{SNR}>8}$~dB, which  consolidate the results from with Fig.\ 4(a) where perfect recovery is attained at  SNR $>8$~dB. The results shown in Fig.\ \ref{fig:mmv+CDI_perfromance}  highlight clearly the advantages of exploiting the prior information about the channel to improve the JUICE performance in terms of activity detection accuracy and channel estimation quality.

\begin{figure}[t]
     \centering
     \begin{subfigure}{0.49\textwidth}
    \includegraphics[scale=.5]{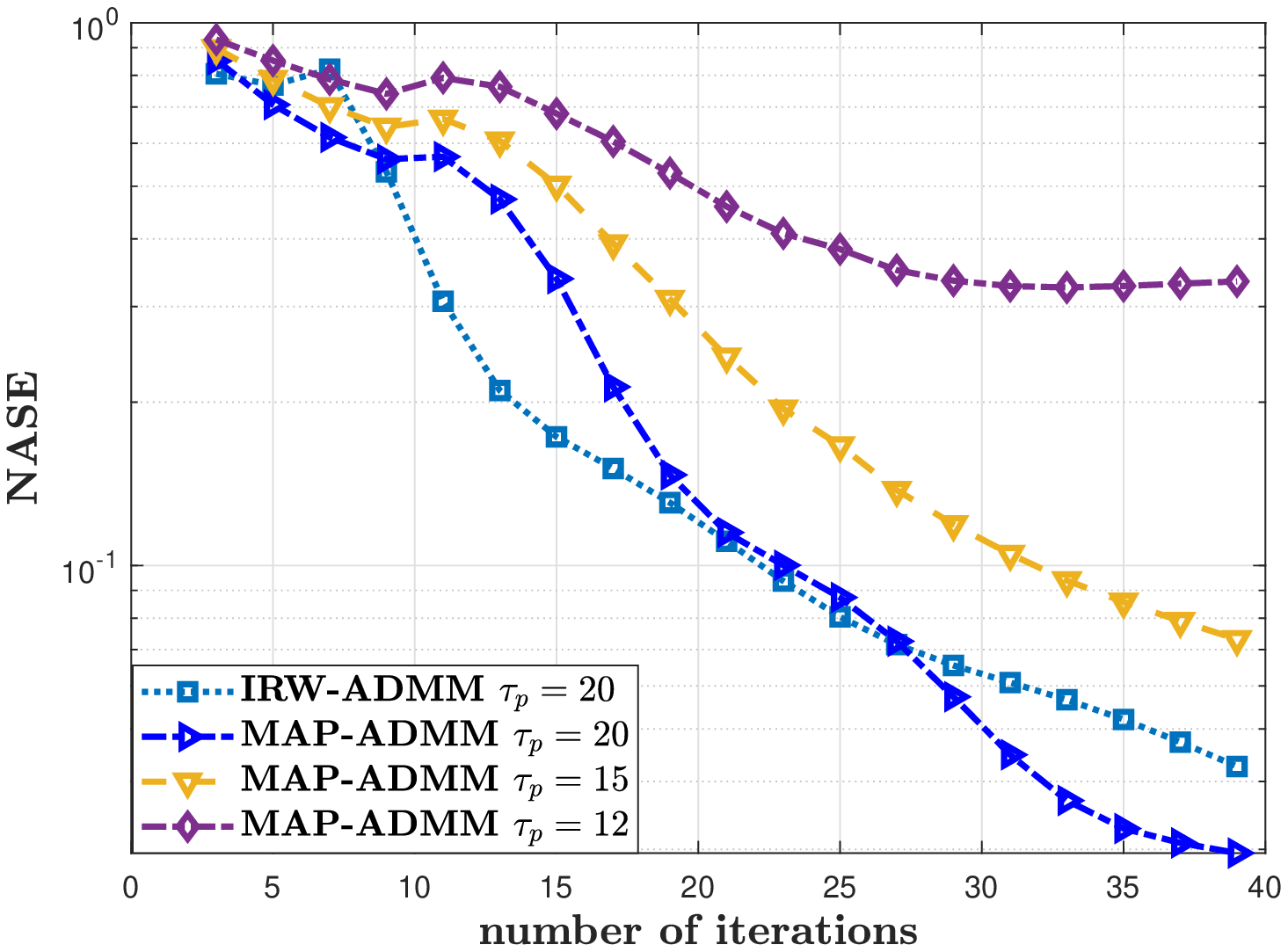}
    \caption{ NASE versus  $(k)$.}
    \label{fig:NASE}
\end{subfigure}
   \hfill      
   \begin{subfigure}{0.49\textwidth}
  \centering
     \includegraphics[scale=.5]{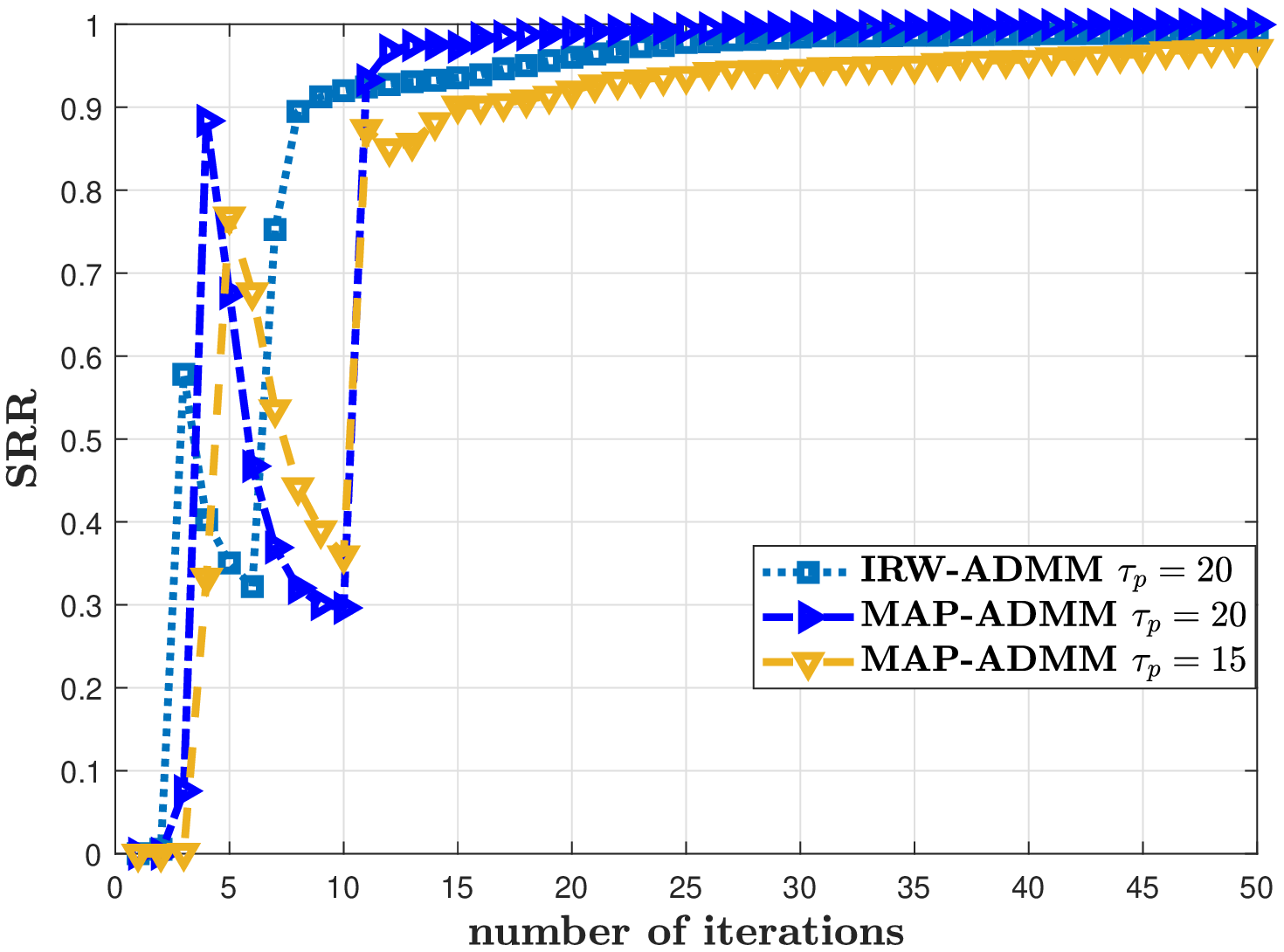}
     \caption{SRR rate versus $(k)$.}
     \label{fig:SR}
\end{subfigure}  
\caption{The performance of the proposed algorithms in terms of NASE and SRR  versus the number of ADMM iterations  for the proposed algorithms  for $N=200$, $M=20$, and $K=10$.}
     \label{fig:convergence_P2}\vspace{-5mm}
 \end{figure}

Fig.\ \ref{fig:convergence_P2}(a) presents a typical convergence behavior of NASE versus the number of ADMM iterations for  MAP-ADMM using different pilot sequence lengths at SNR $=16$~dB. The results reveal that MAP-ADMM using $\tau_{\mathrm{p}}=20$ requires about $40$ iterations to converge, which is similar to IRW-ADMM performance.  On the other hand, the results show that reducing the pilot length  affects also the convergence rate of MAP-ADMM, as it   takes more iterations to converge.

Fig.\ \ref{fig:convergence_P2}(b) plots the SRR performance versus the  average number of ADMM iterations using different pilot lengths. MAP-ADMM using $\tau_{\mathrm{p}}=20$ achieves the perfect activity detection in 20 iterations, whereas it takes up to 40 iterations to achieve the same performance for $\tau_{\mathrm{p}}=15$. This result is interesting as MAP-ADMM needs not to run until convergence in the NASE domain, where it takes up to 40 iterations, rather, MAP-ADMM  can be run for   few iterations until it detects perfectly the set of active UEs (20 iterations on average) as shown  in Fig.\ \ref{fig:convergence_P2}. Afterward, the joint MMSE estimator \eqref{eq::mmse} can be applied on the estimated set of active UEs to provide the optimal  channel estimation quality for the  effective channel matrix, as shown in Fig.\ \ref{fig:mmv+CDI_perfromance}(b).

 \subsection{Effect of the Number of BS Antennas}
\begin{figure*}[t]
    \centering
     \begin{subfigure}{0.48\textwidth}
    \includegraphics[width=1\linewidth]{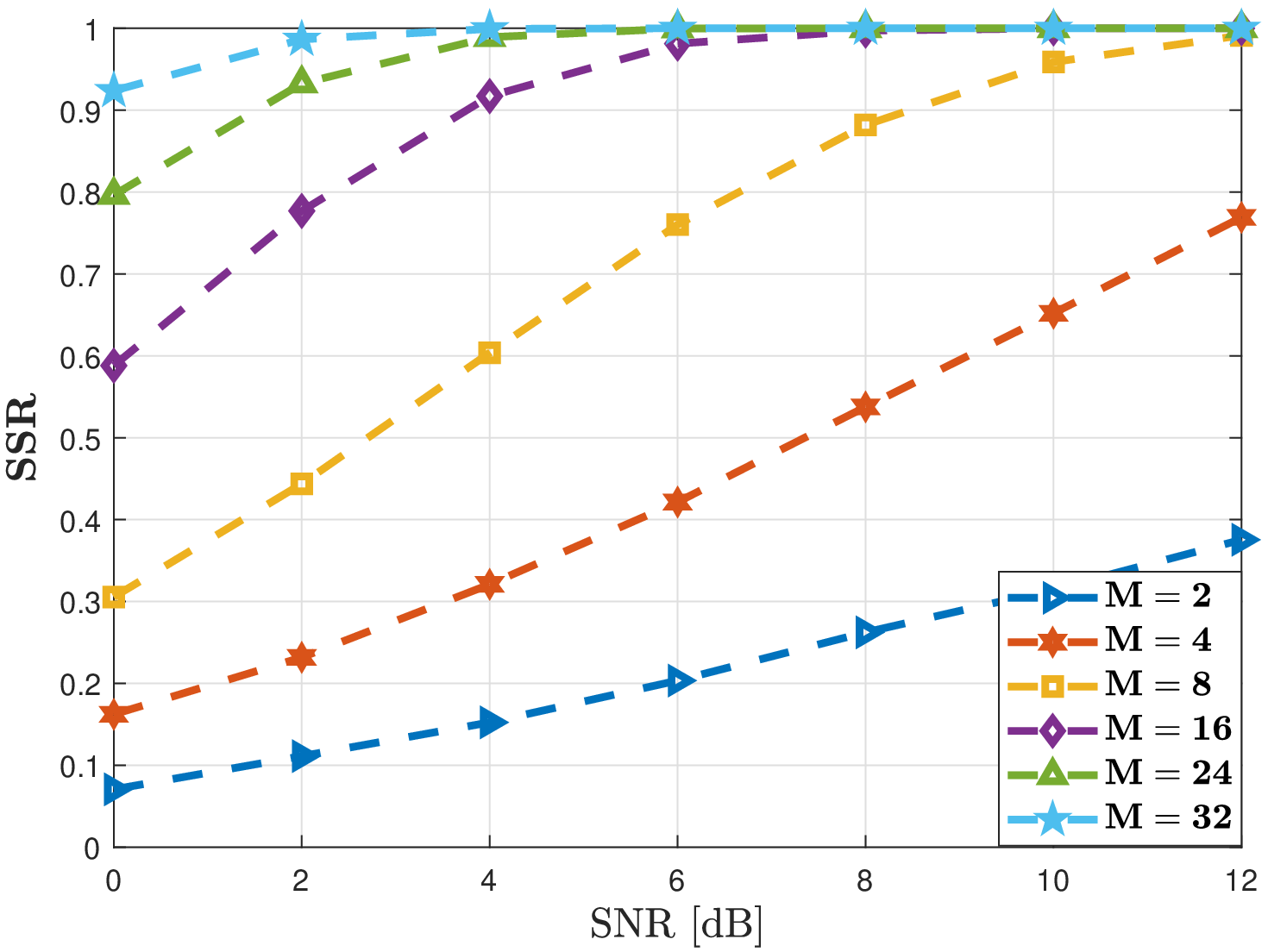}
    \caption{Activity detection.}
    \label{fig:SRR_M}
\end{subfigure}
 \begin{subfigure}{0.48\textwidth}
  \centering
     \includegraphics[width=1\linewidth]{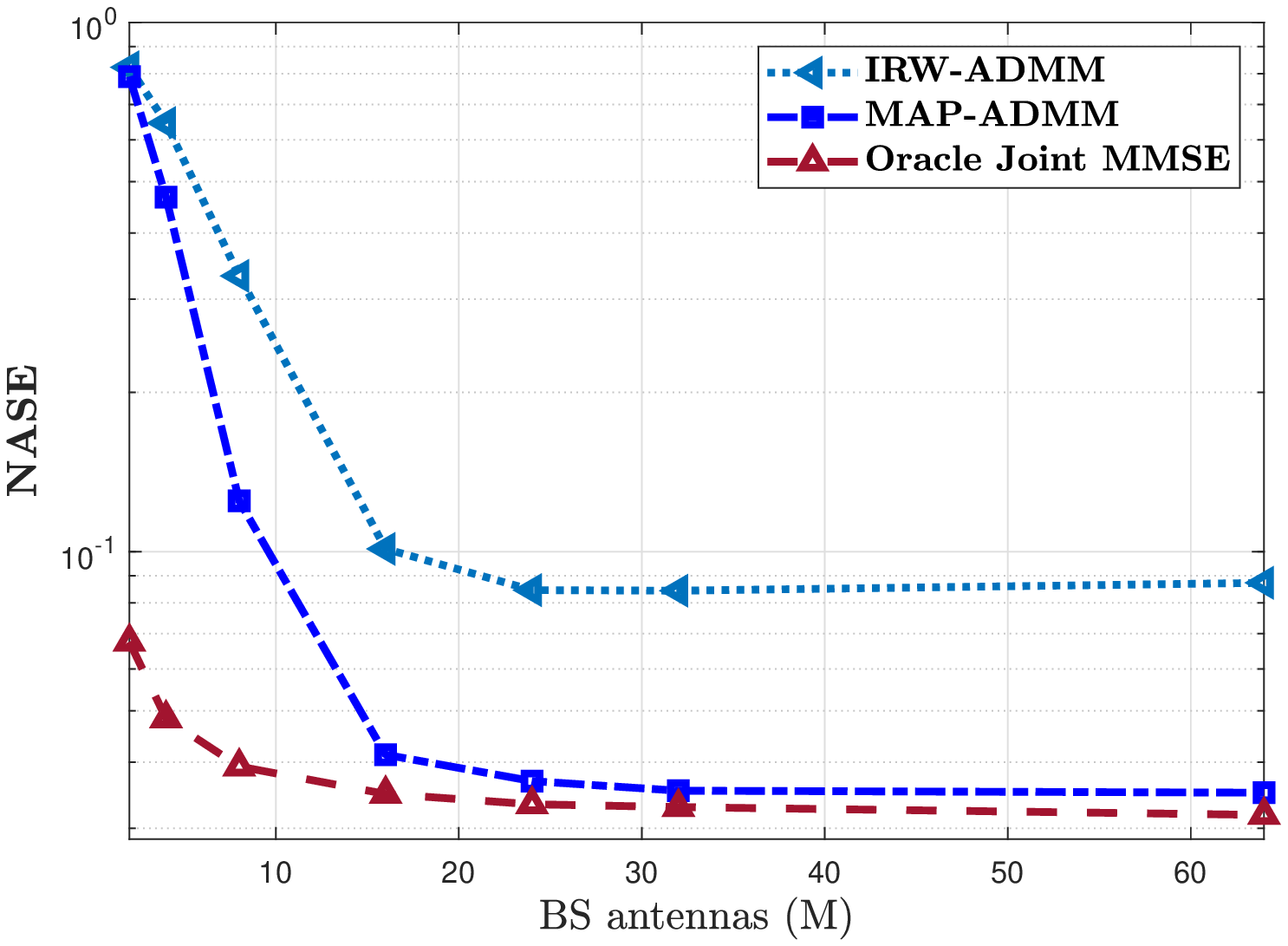}
     \caption{Channel estimation.}
     \label{fig:NASE_M}
\end{subfigure}  
\vspace{-2mm}
\caption{JUICE performance in terms of SRR and NASE versus the  number of BS antennas $M$ for $N=200$, $K=10$,  $\tau_{\mathrm{p}}=20$, and  ${\text{SNR}=16}$~dB.}
\label{fig:M}
\vspace{-5mm}
\end{figure*}

 Next, we focus on quantifying the effect of the number of the BS antennas on the JUICE performance. Fig.\ \ref{fig:M}(a) illustrates the SRR of MAP-ADMM versus
the number of BS antennas $M$. It is clear that increasing the number of BS antennas improves significantly the active user detection accuracy. Moreover, for the low SNR regime, i.e., SNR $<8~$ dB, the results show the significance of increasing the number of antennas to be greater than the number of active UEs, i.e., $M>K$. However, the SRR performance starts to saturate gradually with increasing  the number of BS antennas $M$.  In fact, increasing the number of BS antennas from $M=8$ to  $M=16$ provides more gains than increasing  from $M=24$ to  $M=32$; this means  that the gain  in SRR  gradually decreases as $M$ increases.
Fig.\ \ref{fig:M} (b) depicts the channel estimation performance as a function of the number of BS antennas $M$ at SNR $=12~$ dB. First, as expected,  increasing $M$ improves NASE for all the algorithms.  Moreover, by increasing $M$, the channel estimation quality obtained by MAP-ADMM improves significantly and it approaches the lower bound offered by the oracle joint MMSE. More interestingly, in contrast to activity detection accuracy where the performance saturates when $M>2K$,   channel estimation quality improves  considerably with the increase of $M$.  Fig.\ \ref{fig:M}  points out  the effects of operating in  massive MIMO regime, i.e., $M\!>\!K$: while  user activity detection accuracy saturates around $M\!>\!2K$,  channel estimation quality  consistently  improves when moving to the large numbers of BS antennas $M$.

\vspace{-6mm}
\subsection{Impact of Imperfect Knowledge of the Channel Covariance Matrix}
This section investigates the impact of the training phase to estimate the second-order statistics of the channels $\{\hat{\vec{R}}_i\}_{i=1}^{N}$ on the channel estimation.  More precisely, we vary the number of training samples $T$ and we quantify the  NASE performance of MAP-ADMM and the oracle joint MMSE estimator. Note that once the set of covariance matrices is generated using a particular number of samples $T$, it is used directly as an input to  MAP-ADMM, hence, the BS does not need to update them at each MAP-ADMM iteration.

\begin{figure*}[t]
    \centering
     \begin{subfigure}{0.48\textwidth}
    \includegraphics[width=1\linewidth]{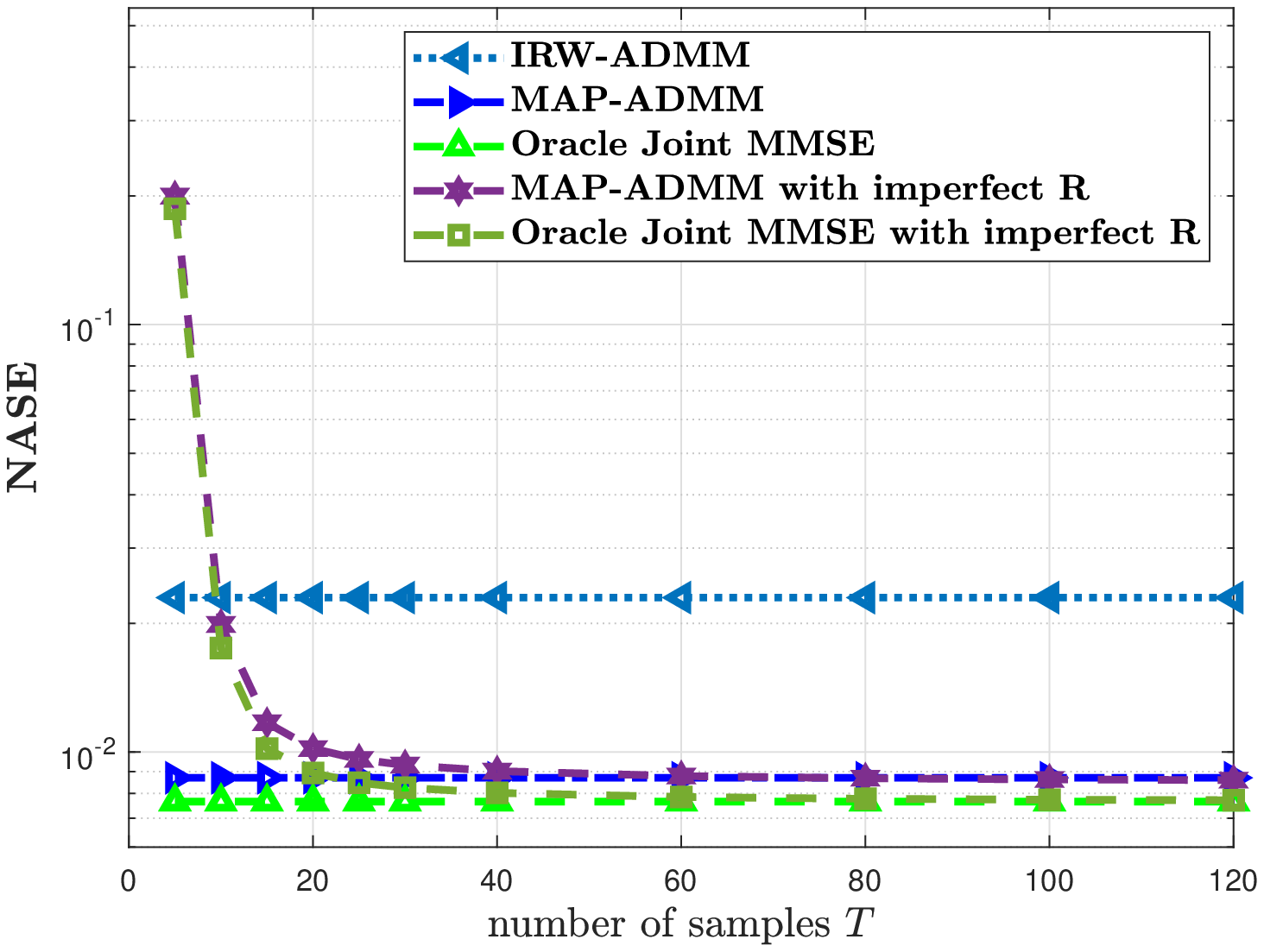}
    \caption{ $M=20$.}
    \label{fig:NASE_T20}
\end{subfigure}
   \hfill      
   \begin{subfigure}{0.48\textwidth}
  \centering
     \includegraphics[width=1\linewidth]{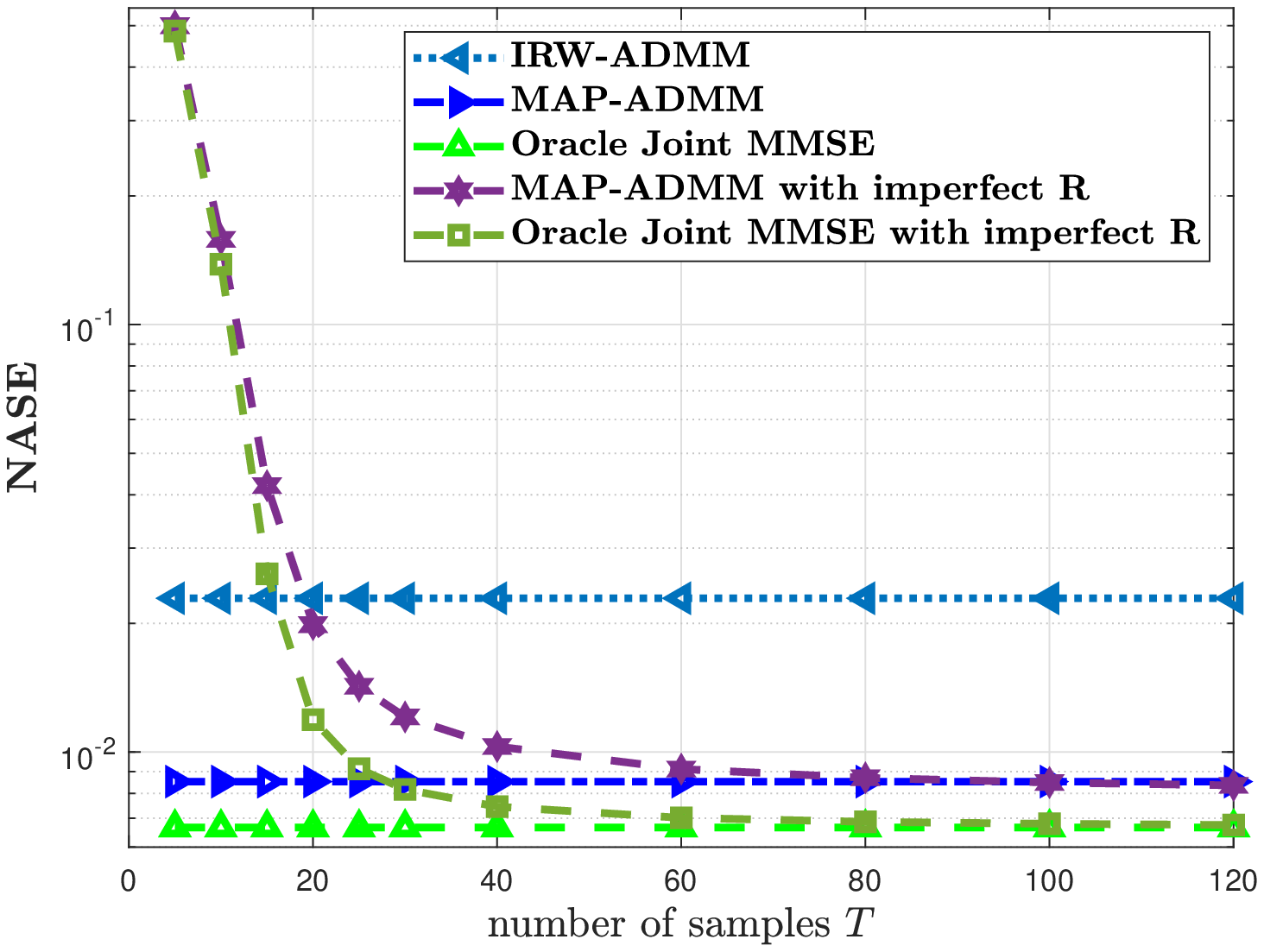}
     \caption{$M=40$.}
     \label{fig:NASE_T40}
\end{subfigure}  
\vspace{-3mm}
\caption{ Channel estimation performance  versus the number of samples $T$, $N=200$, $K=10$, and $\tau_{\mathrm{p}}=20$, ${\text{SNR}=16}$~dB.}
\label{fig:NASE_T}
\vspace{-5mm}
\end{figure*}

Fig.\ \ref{fig:NASE_T} depicts the NASE versus the number of samples $T$ used to generate  $\{\hat{\vec{R}}_i\}_{i=1}^{N}$ for $M=20$ and $M=40$ at ${\text{SNR}=16}$~dB. The regularization parameters for MAP-ADMM and IRW-ADMM  are fixed to the ones providing the best results when perfect knowledge of $\{\hat{\vec{R}}_i\}_{i=1}^{N}$ is available. First, Fig.\ \ref{fig:NASE_T} indicates that using a low number of training samples $T$ is detrimental to the performance of MAP-ADMM  and the joint MMSE estimator  as they require at least $T>\dfrac{M}{2}$ training samples to achieve the same performance as IRW-ADMM. Second, as expected, increasing the number of samples $T$ improves  the channel estimation quality for both MAP-ADMM and the joint MMSE estimator as their NASE asymptotically approaches the lower bounds achieved by their  counterparts that rely on perfect knowledge of $\{\hat{\vec{R}}_i\}_{i=1}^{N}$. More interestingly, the results show that MAP-ADMM and joint MMSE requires around $T=2M$ samples  in order  to a achieve the same NASE  performance to their optimal lower bound.   This results indicate that  MAP-ADMM is not highly sensitive  to  imperfect channel statistics. Finally,  we note that a similar conclusion on the required number samples $T$ to achieve  near-optimal performance for the MMSE estimator is reported in \cite[Sect.~3.3.3]{massivemimobook}.

\section{Conclusions  and Future Work}
\label{conclusion}
The paper addressed the JUICE problem in grant-free access in mMTC under  spatially correlated
fading channels. We presented two JUICE formulations  depending on the availability of CDI. If no CDI is available, we proposed an iterative reweighted $\ell_{2,1}$-norm optimization problem that depends only on the sparsity of the  channel matrix and it is robust and  invariant to different channel distributions. When the CDI is available at the BS, we approached the JUICE from a Bayesian perspective and  proposed a novel JUICE formulation based on MAP estimation. Furthermore,  we derived  ADMM-based algorithms that feature computationally efficient closed-form solutions that can be computed via simple analytical formulas.  

The obtained numerical results highlight the following key findings. 1) Formulating the JUICE as an  iterative reweighted $\ell_{2,1}$-norm minimization problem provides a huge performance improvement over conventional $\ell_{2,1}$-norm minimization. 2) While  incorporating the spatial correlation of the channels  increases the computational complexity  of the recovery algorithms, it  results in significant gains even with a smaller signalling overhead. 3) The performance of the JUICE improve dramatically when moving from the conventional MIMO regime to the massive MIMO regime.  4) The training phase  for estimating the second-order statistics of the channel does not require a substantial amount of resources.   Furthermore, MAP-ADMM is robust  against  imperfect channel statistics knowledge,  which is conducive for practical use cases.
 
MAP-ADMM relies on the knowledge of the CDI at the BS, which may be challenging to acquire in practice. A potential future work is to design a sparse recovery algorithm that estimates the second-order statistics of the channels within the recovery process. Another interesting future direction would be to extend the JUICE framework into multi-cell and cell-free mMTC.

\section*{Appendix}
\subsection{Derivation of $\vec{X}$-update}
First, recall that the  $\vec{X}$-update \eqref{eq::x(k+1)_sec} solves the following optimization problem:
\begin{equation}\label{eq::mix_appendix}
\!\vec{X}^{(k+1)}=\displaystyle\min_{\vec{X}}\sum_{i=1}^{N} \alpha_i^{(k)}\Vert \vec{x}_i \|_{2}+\frac{\rho}{2}\|\vec{X}- \Tilde{\vec{Z}} \|_{\mathrm{F}}^2  +\displaystyle\frac{\rho}{2} \|\vec{X}- \Tilde{\vec{V}}\|_{\mathrm{F}}^2,
\end{equation}
where  $\alpha_i^{(k)}= \beta_1 g_i^{(l)}$, $\Tilde{\vec{Z}}=\vec{Z}^{(k+1)}-\dfrac{\vec{\Lambda}_{\mathrm{z}}^{(k)}}{\rho}$, and $\Tilde{\vec{V}}=\vec{V}^{(k+1)}-\dfrac{\vec{\Lambda}_{\mathrm{v}}^{(k)}}{\rho}$. We can rewrite  \eqref{eq::mix_appendix} as
\begin{equation}
    \begin{array}{ll}
   &  \!\vec{X}^{(k+1)}=\displaystyle\min_{\vec{X}}\sum_{i=1}^{N} \alpha_i^{(k)}\Vert \vec{x}_i \|_{2} +\frac{\rho}{2} \Tr{\bigg(2\vec{X}\vec{X}\herm+\Tilde{\vec{Z}}\Tilde{\vec{Z}}\herm+\Tilde{\vec{V}}\Tilde{\vec{V}}\herm- \vec{X}\herm  \big(\Tilde{\vec{Z}}+\Tilde{\vec{V}}\big) - \big(\Tilde{\vec{Z}}+\Tilde{\vec{V}}\big)\herm \vec{X}}\bigg)\\
         &=\displaystyle\min_{\vec{X}}\sum_{i=1}^{N} \alpha_i^{(k)}\Vert \vec{x}_i \|_{2}+\rho\Tr{\bigg(\vec{X}\vec{X}\herm+\frac{1}{2}\big(\Tilde{\vec{Z}}\Tilde{\vec{Z}}\herm+\Tilde{\vec{V}}\Tilde{\vec{V}}\herm\big)-\frac{1}{2} \vec{X}\herm \big(\Tilde{\vec{Z}}+\Tilde{\vec{V}}\big) - \frac{1}{2}\big(\Tilde{\vec{Z}}+\Tilde{\vec{V}}\big)\herm \vec{X}}\bigg),
    \end{array}\label{eq::X_appendix}
\end{equation}
By subtracting the constant term $\dfrac{\rho}{4}\Vert \Tilde{\vec{V}}-\Tilde{\vec{Z}}\Vert_{\mathrm{F}}^2$ from   \eqref{eq::X_appendix}, and denoting $\vec{S}^{(k)}=\dfrac{1}{2}\big(\Tilde{\vec{V}}+\Tilde{\vec{Z}}\big)$,  the $\vec{X}$-update becomes \eqref{eq::X++}.

\subsection{Joint MMSE Estimator}

The received signal in \eqref{eq::Y} can be rewritten as
 \vspace{-.2cm}
 \begin{equation}
    \vec{y}=\vec{\Theta}_\mathcal{S} \vec{x}_\mathcal{S}+  \vec{w},
    \label{Y_vec}
 \end{equation}
 where $\vec{y}\!=\!\mathrm{vec}(\vec{Y}\tran)\!\in\! \mathbb{C}^{\tau_{\mathrm{p}}M}$, $\!\vec{w}\!\!=\!\!\!\mathrm{vec}(\vec{W}\tran)\! \in \!\mathbb{C}^{ \tau_{\mathrm{p}}M} $, and $\vec{\Theta}_\mathcal{S}=\vecgreek{\Phi}_\mathcal{S}\otimes\vec{I}_{M} \!\in\! \mathbb{C}^{M\tau_{\mathrm{p}} \times KM}$.  The vectorization in \eqref{Y_vec}  transforms the matrix estimation into a classical  vector estimation. Thus, we utilize the MMSE estimator   \cite{kay1993fundamentals}  to \emph{jointly} estimate the  channels of the  active UEs, as 
\begin{equation}
     \vec{x}^{\mathrm{J-MMSE}}=\mathrm{vec}(\vec{X}^{\mathrm{J-MMSE}})=\bar{\vec{x}}+\vec{R}_{\mathrm{diag}}\vec{\Theta}\herm \vec{Q}\big(\vec{y}-\vec{\Theta}\bar{\vec{x}}\big),\label{eq::mmse}
 \end{equation}
where $\vec{Q}=( \vec{\Theta}\vec{R}_{\mathrm{diag}}\vec{\Theta}\herm+\sigma^2 \vec{I}_{\tau_{\mathrm{p}}M} )^{-1}$, $\bar{\vec{x}}$ denotes the mean of $\vec{x}$, and $\vec{R}_{\mathrm{diag}}$ denotes the covariance matrix of  $\vec{x}_\mathcal{S}$ given as a block diagonal matrix whose main-diagonal blocks are given by the scaled covariance matrices $\tilde{\vec{R}}_i$ corresponding to the active UEs $i \in \mathcal{S}$.

\bibliographystyle{IEEEtran}
\begin{spacing}{1.3}
\mybibliography
\end{spacing}
\end{document}